\documentclass[letterpaper,twocolumn,english,aps,pre]{revtex4}
\usepackage{times}
\usepackage[T1]{fontenc}
\usepackage[latin1]{inputenc}
\usepackage{graphicx}

\makeatletter

\newcommand{\NPR}{N_{\mathrm{PR}}}
\newcommand{\nPR}{n_{\mathrm{PR}}}

\newcommand{\gtsimeq}{\raisebox{-0.6ex}{$\,\stackrel
        {\raisebox{-.2ex}{$\textstyle >$}}{\sim}\,$}}

\usepackage{babel}
\makeatother
\begin{document}

\title{Linking physics and algorithms in the random-field Ising model}

\author{Jan H. Meinke}

\author{A. Alan Middleton}

\affiliation{Department of Physics, Syracuse University, Syracuse, NY 13244}

\date{February 15, 2005}

\begin{abstract}
The energy landscape for the random-field Ising model (RFIM) is
complex, yet algorithms such as the push-relabel algorithm exist for
computing the exact ground state of an RFIM sample in time polynomial
in the sample volume. Simulations were carried out to investigate the
scaling properties of the push-relabel algorithm. The time evolution
of the algorithm was studied along with the statistics of an auxiliary
potential field. At very small random fields, the algorithm dynamics
are closely related to the dynamics of two-species annihilation,
consistent with fractal statistics for the distribution of minima in
the potential (``height''). For $d=1,2$, a correlation length
diverging at zero disorder sets a cutoff scale for the magnitude of
the height field; our results are most consistent with a power-law
correction to the exponential scaling of the correlation length with
disorder in $d=2$. Near the ferromagnetic-paramagnetic transition in
$d=3$, the time to find a solution diverges with a dynamic critical
exponent of $z=0.93\pm0.06$ for a priority queue version and
$z=0.43\pm0.06$ for a first-in first-out queue version of the
algorithm. The links between the evolution of auxiliary fields in
algorithmic time and the static physical properties of the RFIM ground
state provide insight into the physics of the RFIM and a better
understanding of how the algorithm functions.
\end{abstract}
\maketitle

\section{Introduction}

Models of materials with quenched disorder, such as spin glasses and
random magnets \cite{YoungBook}, typically have extremely slow dynamics
in the low-temperature glassy phase, due to the existence of many
metastable states separated by barriers that grow with the system
size. Such exponential slowing down affects optimization methods that
are modeled on the dynamics of the physical system, such as simulated
annealing \cite{Kirkpatrick83}. The slow dynamics of the system prevent
precise study of the equilibrium behavior. Besides using physically
motivated dynamics, one can consider using combinatorial methods \cite{Alava2001,HartmannRiegerBook2002}
to compute partition functions or ground states. Some problems, such
as determining the ground state of a 3D spin glass \cite{Barahona}
or finding the partition function for the RFIM at finite temperature,
are NP-hard \cite{JCAd85,SI99}. Though combinatorial approaches derived
in computer science greatly accelerate searches for the ground state
\cite{BarahonaJunger,LiersJungerRinaldiINbook2004}, it is difficult
to study systems with more than $10^{3}$ degrees of freedom. However,
many problems for disordered materials, such as computing the partition
function of a 2D spin glass \cite{Barahona} or finding the ground
state of an RFIM sample \cite{dAuriac}, can be solved in time polynomial
in the volume of the sample, allowing for the solution of samples
with over $10^{7}$ degrees of freedom.

The push-relabel (PR) algorithm introduced by Goldberg and Tarjan
\cite{Goldberg1988} is directly applicable to finding the exact ground
state of the RFIM and has been extensively applied to study the RFIM's
zero-temperature paramagnetic-ferromagnetic transition \cite{BarahonaRFIM,Ogielski1986,HartmannNowak,SourlasEPL,MiddletonFisherRFIM,HartmannRiegerBook2002,HambrickMeinkeMiddleton04,BasteaDuxbury}.
The running time is bound by a polynomial in the number of spins,
but there is a power-law critical slowing down of the PR algorithm
at the zero-temperature ($T=0$) transition \cite{Ogielski1986,MiddletonFisherRFIM,MiddletonCritSlow2002}.

This paper explores the connection between the auxiliary variables
of the PR algorithm and the zero-temperature disorder-driven phase
transition. We also look at the algorithm in the limit of small disorder,
where the dynamics of the algorithm turn out to be closely related
to annihilation processes studied in statistical physics \cite{ADR98,FLSR92,toussaint:2642}.

We define the RFIM Hamiltonian, review its phases, and define the
rules for the push-relabel dynamics in Sec.~\ref{sec:Random-field-Ising}.
These rules describe the evolution of auxiliary fields; the dynamics
of these fields leads directly to a ground state for the RFIM. Our
main focus will be on the {}``height'' or potential field. This
field guides the determination of the spin-up and spin-down domains
in the ground state. Given these definitions, we outline our results
in Sec.~\ref{sec:Overview}. We study the relationship of the PR
algorithm to annihilation processes at low values of the disorder
by studying the algorithm dynamics and final topography of this auxiliary
potential surface in Sec.~\ref{sec:Small-disorder-limit}. In Sec.~\ref{sec:General-disorder-and},
we study the topography of the potential surface for general disorder,
especially near the paramagnetic-ferromagnetic transition. The primary
results of the paper, especially the scaling of the running time and
statistics of the potential field, are summarized in Sec.~\ref{sec:Summary}.

\section{Random-field Ising model and the push-relabel algorithm\label{sec:Random-field-Ising}}

The random-field Ising model (see, e.g., \cite{YoungBook} and references
therein) captures essential features of models in statistical physics
that are controlled by disorder and have {}``frustration'', i.e.,
energy competition between different terms of the Hamiltonian. Such
systems have complex energy landscapes and the ground-state configuration
for a given sample is not usually obvious. Large barriers separate
a very large number of metastable states. If such models are studied
using simulations mimicking the local dynamics of physical processes,
it takes an extremely long time (exponential in a power of the system
size) to encounter the exact ground state. But in many cases there
are very efficient methods for finding the ground state. These methods
break away from a direct physical representation. Extra degrees of
freedom are introduced and an expanded problem is solved. By expanding
the configuration space and choosing proper dynamics, the algorithm
goes {}``downhill'' in a fashion that avoids having to go over barriers
that exist in the original physical configuration space. An attractor
state in the extended space is found in time polynomial in the size
of the system. When the algorithm is completed by finding this attractor
or minimum in the extended space, the auxiliary fields can be projected
onto a physical configuration, which is the guaranteed ground state.
The RFIM is an example where this extension can be carried out.

\subsection{Model}

In the RFIM, there is competition between ferromagnetic terms, characterized
by a strength $J$ and local random fields $h_{i}$ of characteristic
magnitude $J\Delta$, with a Hamiltonian \begin{equation}
\mathcal{H}=-J\sum_{\langle ij\rangle}s_{i}s_{j}-\sum_{i}h_{i}s_{i},\label{eq:RFIM_Hamiltonian}\end{equation}
 where the Ising spins $s_{i}=\pm1$ lie on sites $i$ on a $d$-dimensional
lattice and the notation $\left<ij\right>$ indicates a sum over nearest-neighbor
pairs of sites. The samples have linear dimension $L$, with $n=L^{d}$
sites, and we use periodic boundary conditions. The independent Gaussian
random variables $h_{i}$ have mean $0$ and variance $\Delta^{2}J^{2}$.

In dimensions $d>2$, there is a zero-temperature transition between
two phases at the critical disorder $\Delta=\Delta_{c}$. When $\Delta<\Delta_{c}$,
the ferromagnetic interaction between nearest neighbors dominates
and the spins take on a mean value $m=n^{-1}\sum_{i}s_{i}$ with $\left|m\right|\neq0$
in the limit $n\rightarrow\infty$. In the case $\Delta>\Delta_{c}$,
randomness dominates and the ground state is {}``paramagnetic'',
with $\left|m\right|=0$, as $n\rightarrow\infty$. In the standard
picture, the zero-temperature transition has the same critical exponents
as the finite temperature transition.

In dimensions $d=1,2$, there is no zero-temperature transition. For
a given sample size $L$, there is a characteristic crossover value
$\Delta_{x}(L)$ for the disorder strength. When $\Delta\ll\Delta_{x}(L)$,
samples of size $L$ have a high probability to have uniform spin.
For large disorder, $\Delta\gg\Delta_{x}(L)$, there are many domains
of uniform spin in the ground state. Exact calculations and scaling
arguments show that $\Delta_{x}(L)\sim L^{-1/2}$ for $d=1$, while
scaling arguments and computer simulations give $\Delta_{x}(L)\sim[\ln(L)]^{-1/2}$
for $d=2$, up to logarithmic corrections \cite{ImryMa1975,GrinsteinMa82,Nattermann1988,NattermannRFIMreview,SeppalaPetajaAlava2D1998,SeppalaAlava2D2001,SchroderKnetterAlavaRieger2002}.

\subsection{\label{sec:Generic-Push-Relabel-Algorithm.}Push-Relabel Algorithm}

We present here a description of the auxiliary fields and algorithmic
dynamics used in the push-relabel algorithm. We do not provide the
standard proof of the correctness of the algorithm (see \cite{Alava2001,HartmannRiegerBook2002,Middleton2004,HambrickMeinkeMiddleton04}
for such proofs and further discussion). To describe our results,
it is necessary to define the auxiliary variables used and the update
rules. It is also useful to provide an intuitive description of the
dynamics of the algorithm.

There are three auxiliary fields: (i) the excess $e_{i}$, (ii) the
residual strength (capacity) $r_{ij}$ defined for ferromagnetically-coupled
pairs $\left<ij\right>$, and (iii) a distance or height field $u_{i}$.
Initially, the fields are set according to the rules $e_{i}=h_{i}$,
$r_{ij}=J$ and $u_{i}=0$. A site $i$ is {}``active'' if $e_{i}>0$
and $u_{i}<\infty$; a site is a {}``sink'' if $e_{i}<0$. Primarily
two types of operations, {}``push'' and {}``relabel'', modify
the fields at active sites and their neighbors. The push operation
rearranges the locally conserved $e_{i}$ and also modifies the $r_{ij}$.
The conditions for a push from site $i$ to neighboring site $j$
are that $u_{i}=u_{j}+1$, $e_{i}>0$ and $r_{ij}>0$. If a push is
executed, the excess $e_{i}$ is modified via $e_{i}\rightarrow e_{i}-\delta_{i}$,
where $\delta_{i}=\min(e_{i},r_{ij})$, and the residual bond strengths
are modified according to $r_{ij}\rightarrow r_{ij}-\delta_{i}$,
and $r_{ji}\rightarrow r_{ji}+\delta_{i}$. The relabel operation
increases the $u_{i}$ at an active site $i$. Whenever a push at
an active site $i$ is not possible, $u_{i}$ is increased to the
minimum value to enable a push from $i$ ($u_{i}$ is set $u_{i}=\infty$,
if no push is possible from $i$ due to saturated bonds). As seen
from the rules for a push, the height field $u_{i}$ guides the rearrangement
of excess. The pushes are always downhill with respect to the height
field.

As a motivation, though again, not a proof, pushing the excess field
roughly corresponds to a rearrangement of the external magnetic fields.
Regions that have large ferromagnetic coupling compared to $\Delta$
will have a net excess that is of the same sign as the total magnetic
field over the region. The pushes allow the spin orientations in these
domains to be determined by cancellation of positive and negative
excess. Note that as pushes are allowed only when $r_{ij}>0$, the
rearrangement of excess is limited. This leads to finite size domains,
for large enough samples and strong enough disorder. 

The set of rules given so far does not define an algorithm. In order
to have a well-defined procedure, one must organize the push and relabel
operations and set a criteria for termination. When considering a
site with $e_{i}>0$, we first execute all possible pushes and then
relabel site $i$, if necessary. This is defined as a single push-relabel
step; the number of such steps will be our measure of algorithmic
time. The order in which sites are considered is given by a queue.
In this paper, we compare results for two types of queues: a first-in-first-out
queue (FIFO) and a lowest height priority queue (LPQ) (other options
are available \cite{CherkasskyGoldberg97,Seppala1996,HambrickMeinkeMiddleton04}).
The FIFO structure executes a PR step for the site $i$ at the front
of a list. If any neighboring site is made active by the PR step,
it is added to the end of the list. If $i$ is still active after
the PR step, it is also added to the end of the list. This structure
maintains and cycles through the set of active sites. The LPQ structure
is also a list, but is always sorted by the height label $u_{i}$.
The sites with lowest $u_{i}$ are always at the front of the list.
The LPQ version executes PR steps for active sites that are likely
to be near sinks. This algorithm will act repeatedly on the set of
sites with lowest height. 

The algorithm terminates when no active sites remain. Sites with positive
excess will remain, in general, but the height at those sites will
be $u_{i}=\infty$. At the end of the algorithm, the sign of spin
$s_{i}$ is set to be positive if there is no path from $i$ to a
site with negative excess along bonds $\left<uv\right>$ with $r_{uv}>0$.
If there is such a path, the spin $s_{i}=-1$ in the ground state. 

One operation that greatly speeds up the running time of the algorithm
is the global update \cite{Goldberg1988,HartmannRiegerBook2002,Middleton2004,HambrickMeinkeMiddleton04}.
This operation recomputes the height field $u_{i}$ so that $u_{i}$
is the minimal distance from the site $i$ to the set of sinks. The
height is set to $u_{i}=\infty$ when there is no path from $i$ to
a sink along with all edges satisfying $r_{uv}>0$. We denote the
period between global updates as $\Gamma$. In the rest of this paper,
we fix $\Gamma=n$ ($d=2,3$) or $\Gamma=2n$ ($d=1$), which gives
near minimal running times for the PR algorithm with fixed global
update intervals \cite{Seppala1996,HambrickMeinkeMiddleton04}.

\section{Overview of results\label{sec:Overview}}

Our principle results concern the statistics of the height field $u_{i}$
and the connection between the topography and the running time of
the ground state algorithm. The local relabels and global updates
result in a height field that guides positive excess to maximal cancellation
with the negative excess sinks, subject to constraints from the residual
bond strengths $r_{ij}$. The final configuration of the height field
$u_{i}$ determines the ground state, though a single ground state
can be consistent with a large set of terminal height field configurations.
The choice of the order of operations will determine the final height
field in a given sample. The time scale needed to establish this height
field, while cancelling excess fields and modifying residual bond
strengths, determines the running time of the algorithm. In this paper,
we mostly study the FIFO data structure. Besides being the fastest
data structure we used, this structure is most natural for making
connections to physical dynamics, where particles move in parallel.
For this structure and algorithm, sites with positive excess are each
treated once during a cycle through the active sites. This is to be
contrasted with a data structure like LPQ, where a single parcel of
excess may be moved many times while other portions of the system
remain static.

Using the FIFO data structure, we first study the limit of small $\Delta$,
where the rearrangement of excess is not affected by the bond strength.
By rearranging the positive excess, the algorithm cancels out negative
and positive excess as much as possible. We arrive at the natural
connection that the dynamics of the algorithm at weak fields is related
to the extensively studied set of annihilation models $A+B\rightarrow\emptyset$.
In such models, there are two types of particles, $A$ and $B$, with
one or both types mobile. The particles annihilate (or combine to
an inert particle) upon contact. In general, the motion of the particles
is modeled as due to random diffusion or to overdamped drift caused
by interactions between the particles. A very rich set of scaling
results have been found to describe the dynamics of the average density,
domain sizes, and domain profiles in the annihilation process \cite{ADR98,AVG97,FLSR92,odor:663,toussaint:2642}.
For a description of the push-relabel algorithm, we can consider positive
excess as $A$ particles and the sinks as $B$ particles. The $B$
particles are immobile. For Gaussian disorder, the particles will
not exactly annihilate: upon meeting, either the positive excess or
negative excess will be saturated by the excess of opposite sign.
However, the cancellation of negative and positive excess leads to
a decrease in the density of active sites similar to the direct cancellation
$A+B\rightarrow\emptyset$. Particles of type $A$ may coalesce, changing
the speed at which the densities evolve. In the small disorder limit,
we find that the running time grows very slowly - apparently logarithmically
in $L$ - when measured in the number of PR steps per site. The final
distribution of sinks is found to have a fractal character at small
length scales; this fractal character is consistent with that for
annihilation processes \cite{FLSR92}, at least when $d=1$. This
fractal distribution is related to a power-law distribution for the
height values. We also study the time evolution for the density of
$A$ and $B$ particles. When the $A$ particles can coalesce, the
result is a single excess packet of large weight at long times. Forbidding
this coalescence by modifying the algorithm leads to an approximate
equal density of sinks and active packets during the solution process.

We also studied the number $\NPR$ of push-relabel operations required
to find the ground state and the topography of the height field $u_{i}$
for general $\Delta$. For $d=1,2$, the peak running times were used
to define the crossover field $\Delta_{x}$. The size dependence of
$\Delta_{x}$ is consistent with the expectations $\Delta_{x}\sim L^{-1/2}$
($d=1$) and $L\sim\left(\Delta_{x}\right)^{-2y}e^{-\Delta_{0}^{2}/\Delta_{x}^{2}}$,
with $y\approx1$ and a fitted value for $\Delta_{0}^{2}\approx1.3$.
The scaling of the height fields in $d=1,2$ is consistent with this
same divergence in correlation length $\xi$: the fraction of sites
$P(u)$ with height $u$ is proportional to a function of $u/\xi$.

For $d=3$, the running times near $\Delta_{c}$ exhibit distinct
dynamic critical exponents for the running times. Excellent scaling
results when we use values for the critical disorder $\Delta_{c}$
and correlation length exponent $\nu$ derived from more physical
measures of the RFIM ground state \cite{MiddletonFisherRFIM}. For
LPQ we find the dynamical critical exponent $z=0.93\pm0.06$, while
for FIFO, $z=0.43\pm0.06$, where $z$ describes the running time
via $\NPR\sim L^{z}$. The probability distribution for height fields
decreases exponentially with height $u$ when $\Delta>\Delta_{c}$,
while it is increasing at small $u$ for $\Delta<\Delta_{c}$. At
the critical point $\Delta=\Delta_{c}$, the probability distribution
$P(u)$ is very nearly constant out to the linear size of the system.
We also study the structure of the domains by analyzing the paths
to the sinks near $\Delta_{c}$; these paths are apparently nonfractal
for all $\Delta$.

\section{\label{sec:Small-disorder-limit}Small disorder limit}

The relationship between the auxiliary fields and running times is
simplest when the random field is weak compared to the magnitude of
the exchange coupling. In this limit, the residual capacities of the
directed links connecting sites are never saturated by pushes and,
as a result, the push operations are unrestricted. The dynamics, then,
is roughly described by the motion of positive excess towards the
sinks. This would be exactly true if global updates were carried out
immediately whenever a sink was removed by annihilation with positive
excess. The height field would then always guide positive excess directly
towards the nearest negative sinks. Note that to maintain this exact
guidance one need do even less: upon annihilation of a sink, the height
field only in the region that acted as a funnel for that sink would
need to be updated. We restrict ourselves to the standard approach
using only local relabels and global updates at periodic intervals.

For much of the evolution time of the algorithm, the global updates
maintains a $u_{i}$ landscape that approximates well the one that
would be found for more rapid updates (at least at early times, when
there are many active sites). We note that for very weak disorder,
where residual bonds are never saturated, a global update constructs
a height field that exactly equals a potential field where the negative-excess
sites (the sinks) act as sources. This potential is the minimum over
all sinks of a potential that increases linearly with distance from
a sink. The packets of positive excess at active sites are guided
by this potential from the sinks but do not interact with each other
via any potential. This lack of repulsion between the $A$ particles
is one difference from the force-guided motion for annihilation processes
that has been previously considered \cite{ADR98}. As already noted,
the positive excess packets do interact by coalescence when an excess
is pushed onto a site already containing excess.

Numerically, we studied the low disorder limit using two varieties
of the algorithm. We first varied $\Delta$ and examined the $\Delta\rightarrow0$
limit. To determine what disorder parameters gave this limit, we examined
a wide range of values with $\Delta\ll1$. We used this data to determine
a value of $\Delta$ where the quantity of interest (such as $N_{PR}$)
was constant over a factor of at least $10^{3}$ in $\Delta.$ This
assured that the small-disorder limit had been reached, but that $\Delta$
was not so small that the discreteness of the disorder distribution
affected the running time. In our algorithm, integer values for $J$
and $h_{i}$ were used, with $J=5\times10^{8}$. The product $J\Delta$
characterizes the integer resolution for the disorder; small values
do affect the results. For example, a small decrease in $\NPR$ was
seen for $J\Delta<10^{2}$. In samples of less than $2\times10^{6}$
spins, we found a range of $\Delta$ where the running time and other
quantities were quite constant. For the running time data reported
here, where finite $J$ is used, a value of $J\Delta$ of value $10^{3}$
or $10^{4}$ was well within this range. Some quantities, such as
the number of sites with positive excess at the end of the algorithm,
required even smaller values of $\Delta$ at fixed $J$, in the largest
samples. This may be somewhat surprising, as the magnitude of the
excesses that are rearranged is still always much less than $J$,
so that a single packet of positive excess will not saturate a bond.
However, some bonds end up being on the path of many packets, so that
their residual bond strength is driven towards zero. Because of this
detail, we compared our results against a second program in which
$J=\infty$. In this version of the algorithm, there were no limits
set on the capacity of a bond. Besides allowing one to study the $\Delta\rightarrow0$
limit directly, this code is also simpler than the full push-relabel
code and requires less memory, as we do not need to maintain the $2d$
values of $r_{ij}$ at each site, allowing us to study samples up
to size $512^{3}$. We verified the results of this simpler code on
a sample-by-sample basis in many cases to verify that the results
agreed with the $\Delta\rightarrow0$ limit of the full push-relabel
program which did not take $J\rightarrow\infty$. After confirming
the correctness of the newer approach, we used it to generate most
of the data used in this subsection.

\subsection{\label{sub:TimeSmallDelta123}Running time at small $\Delta$}

\begin{figure}
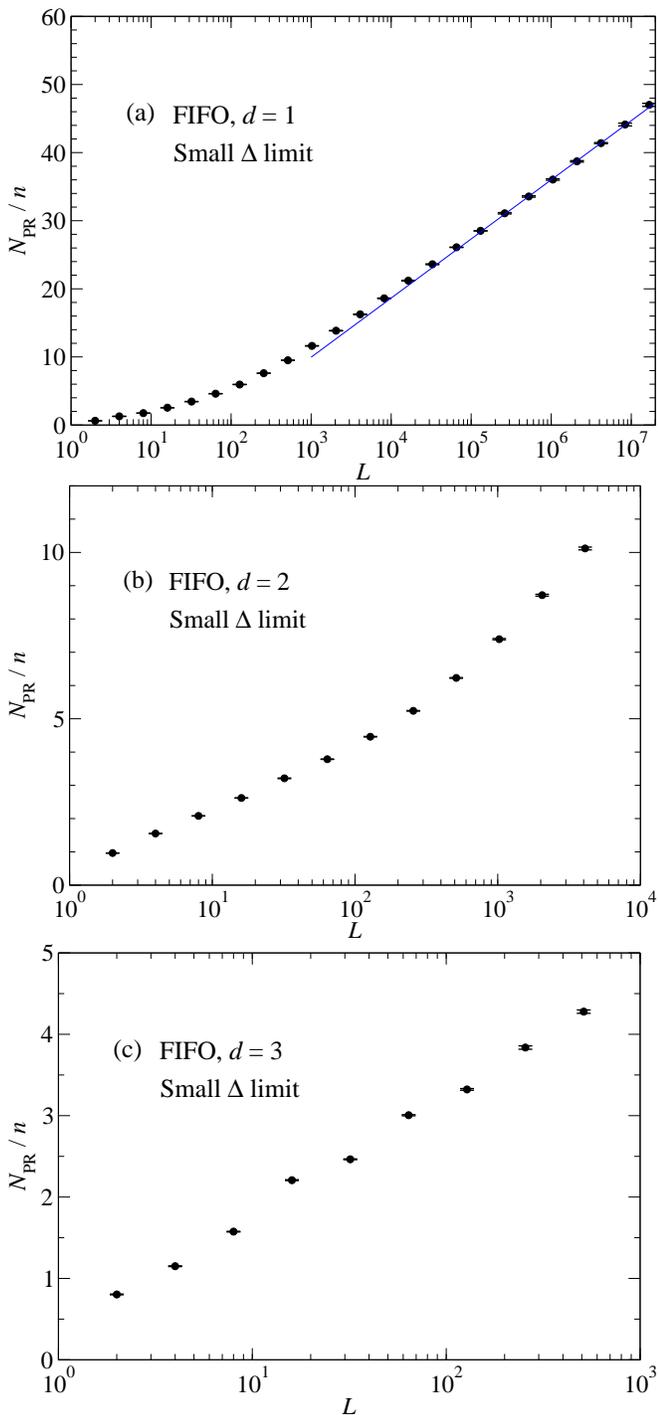

\includegraphics[%
  width=1.0\columnwidth]{Fig1a.eps}

\includegraphics[%
  width=1.0\columnwidth]{Fig1b.eps}

\includegraphics[%
  width=1.0\columnwidth]{Fig1c.eps}

\caption{\label{cap:NPR_weakdisorder}Dependence of the running time, measured
by the number of push-relabel steps per spin, $\NPR/n$, on the system
size, in the limit of small disorder $\Delta$, for dimensions (a)
$d=1$ ($L=2\rightarrow2^{24}$), (b) $d=2$ ($L=2\rightarrow4096$),
and (c) $d=3$ ($L=2\rightarrow512$). The FIFO data structure was
used in each case. The error bars indicate the $1\sigma$ error in
the statistic; the minimum number of samples at the largest sizes
is $1200$, $3700$, and $876$ for $d=1,$ 2, and 3, respectively,
with more samples at smaller $L$. The dependence is suggestive of
a logarithmic dependence $\NPR\sim L^{d}\ln(L)$, especially when
$d=1$, as indicated by the fit in (a) for $2^{24}\ge L\ge2^{18}$.}
\end{figure}

A direct way to measure the dynamics of the algorithm is to examine
the dependence of the running time, measured by the number of push-relabel
operations $\NPR$, on system size $L$. We first present such data
for the case of very weak disorder. The dependence of $\NPR/n=\NPR L^{-d}$
on $L$ is plotted for $d=1,2,3$ in Fig.~\ref{cap:NPR_weakdisorder}.
For the 1D systems, we studied samples over the size range $L=2\rightarrow2^{24}$.
The data is consistent with an asymptotic approach to $L^{-1}\NPR\sim\ln(L)$,
though the apparent slope of the $L^{-1}\NPR$ vs.~$L$ plot becomes
approximately constant only for larger $L\gtsimeq5\times10^{3}$.
For the last two decades in scale, a logarithmic fit describes the
data well. The growth in $\NPR$ with $L$ for $d=2$ is also very
slow compared to all but the smallest power laws. The slope on a linear-log
plot of $\NPR/n$ vs.~$L$ is not constant over the length scales
studied, but the data is not inconsistent with convergence to $\NPR\sim L^{2}\ln(L)$
for $L$ larger than a crossover point $L_{x}\sim10^{3}$, similar
to the $d=1$ crossover in shape and at roughly the same $\NPR/n$
or $L$. The 3D data is also consistent with very slow growth for
the running time at small $\Delta$, but given the $d=1$ and $d=2$
results and that we are restricted to smaller $L\le512$ and $\NPR/n<5$,
it is not possible to make a strong conclusion on a functional form
for $\NPR(L,\Delta\rightarrow0)$ when $d=3$. Over the size range
that we studied in the case $d=3$, the slope is not smoothly varying
on a linear-logarithmic plot of $\NPR/n$ vs.~$L$, due to the discreteness
resulting from global updates seen at small $\NPR/n$, but the approximate
behavior is logarithmic over this range of scales.

A plausibility argument can be made for a logarithmic dependence of
$\NPR/n$ on $L$. If the algorithm successively {}``solves'' for
the ground state by coalescing positive and negative excesses, unhindered
by residual capacity constraints, one might expect that the coalescence
is carried out on successively larger scales, leading to a $\ln(L)$
number of intermediate solutions. The pushes lead to cancellation
of excess on a given scale. If each length scale requires a constant
number of push-relabel operations per site, this would give $\NPR\sim L^{d}\ln(L)$.
In $d=1$, this is the most likely scenario, as rearranging excess
over a scale $\ell$ requires a minimum of $\ell$ PR steps. In higher
dimensions, however, not all sites must be activated in a volume $\ell^{d}$
for the positive excess to rearrange on a scale $\ell$. A better
understanding of the dynamics than is presented here is needed to
confirm this tentative description in higher dimensions. To explore
the dynamics further, we next consider how many sites with positive
excess are being rearranged and the number of sinks present.

\subsection{\label{sub:Number-of-remnant}Number of remnant sinks and sources}

The ground-state magnetization is uniform but has random sign in a
finite system in the limit of weak disorder. The topography of the
height field, however, is very different between the two possible
ground states (up and down). If the spins are all positive at the
completion of the push-relabel algorithm, there are no sites with
negative excess, and $u_{i}=\infty$ at all $i$, as all sites are
{}``unreachable''. If the magnetization is uniformly negative, there
are no sites with positive excess, $u_{i}$ is finite at all $i$,
and there is a remnant set of negative excess sites, i.e., the sinks
that have survived annihilation up to the completion of the PR algorithm.
The number of sites with positive and negative excess gives some indication
of the dynamics of the algorithm.

\begin{figure}
\includegraphics[%
  width=1.0\columnwidth]{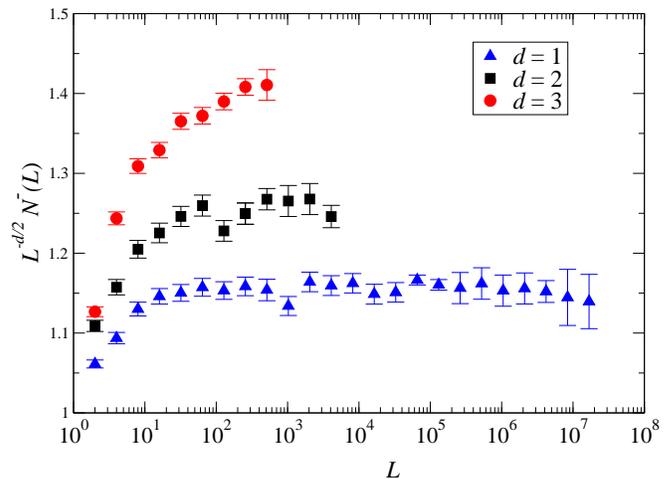}

\caption{\label{cap:Nminus}{[}Color online{]} Plot of $L^{-d/2}N^{-}(L)$,
the average number of sites with negative excess at the termination
of the algorithm scaled by the expected average $L^{d/2}$, vs.~system
size $L$, for $d=1,2,3$ samples with negative magnetization, in
the $\Delta\rightarrow0$ limit. The plot indicates convergence to
a single value for the scaled variable in each dimension, consistent
with $N^{-}\sim L^{d/2}$.}
\end{figure}

As the $h_{i}$ are independent variables with identical distribution,
the magnitude of the sum of the $h_{i}$ scales simply as $\sim L^{d/2}J\Delta$.
In the FIFO and LPQ algorithms, the negative excess is not mobile
and so cannot coalesce. If the mean negative excess at a remnant sink
at the end of the algorithm is of order -$J\Delta$, as it is at the
beginning of the algorithm, we expect that $\sim L^{d/2}$ sites with
negative excess remain. This is the most natural assumption, taking
the cancellation of negative excess to be with packets of positive
excess that are either comparable or much larger in magnitude. To
verify this assumption, we may simply count the number of sites with
negative excess at the end of the algorithm in samples with negative
magnetization. We will refer to the average of this quantity as $N^{-}(\Delta,L)$.
Plots of $L^{-d/2}N^{-}(\Delta=0,L)\equiv L^{-d/2}N^{-}(L)$ for $d=1,2,3$
are displayed in Fig.~\ref{cap:Nminus}. In all dimensions studied,
there is a convergence to a single value. This is consistent with
the picture that the mean negative excess at a sink converges to a
single value $\sim-J\Delta$ as $L\rightarrow\infty$.

\begin{figure}
\includegraphics[%
  width=1.0\columnwidth]{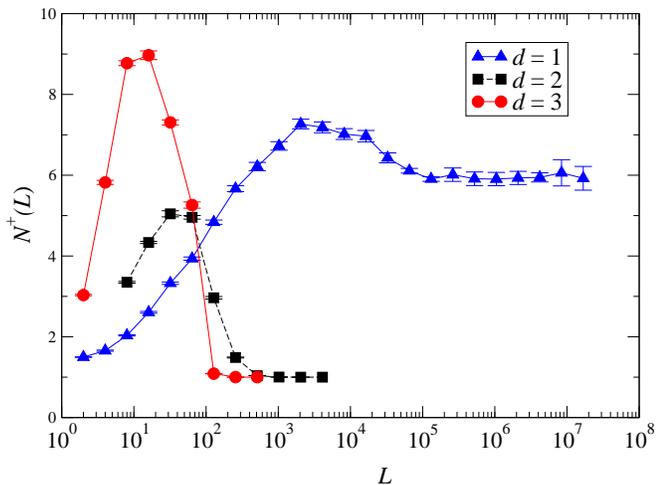}

\caption{\label{cap:Nplus}{[}Color online{]} Plot of $N^{+}(L)$, the number
of remnant sites with positive excess in samples with positive magnetization,
vs.~$L$ for $d=1,2,3$, in the limit $\Delta\rightarrow0$. The
small numbers indicate that most of the positive excess is collected
into less than 10 sites by the termination of the algorithm. The 1D
data appear to converge to a constant $N^{+}(\infty)=5.96\pm0.03$,
while the 2D and 3D data apparently converge to $N^{+}(\infty)=1$ }
\end{figure}

In contrast, for the samples with positive magnetization in the ground
state, the number of sites $N^{+}(L)$ with positive excess does not
vary rapidly with $L$. The number of packets of positive excess at
the termination of the algorithm is generally small, with the average
of $N^{+}(L)<10$ in all sizes $L$ and dimension $d$ that we examined
(see Fig.~\ref{cap:Nplus}). As the net positive excess is the sum
of $h_{i}$ in these samples, the amount of excess per site is large,
$\approx O(L^{d/2}\Delta)$. The initial excesses of magnitude $\Delta$
coalesce significantly under the push operations at small $\Delta$.
At initial times and high density of sites with positive excess, this
coalescence may take place due to {}``chance'' collisions that depend
on the order of the operations at the sites. At lower densities, the
coalescence must result from a focusing caused by the topography of
the height landscape $u_{i}$.

\begin{figure}
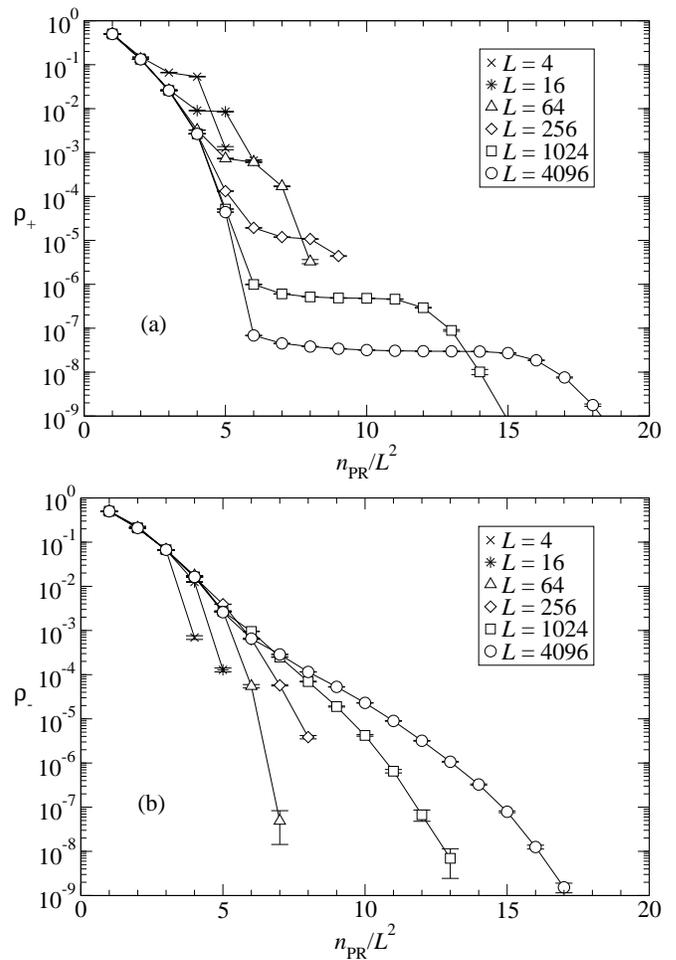

\includegraphics[%
  width=1.0\columnwidth]{Fig4a.eps}

\includegraphics[%
  width=1.0\columnwidth]{Fig4b.eps}

\caption{\label{cap:DensityVsTime2D}Plots of the number densities $\rho^{+}$
and $\rho^{-}$ vs.~$\nPR$, the number of PR steps executed during
the algorithm for (a) positive excess packets (active sites) and (b)
sinks, for $d=2$ FIFO, in the limit of small disorder. The density
of positive sinks decreases very rapidly until, in the larger systems,
a plateau $\rho^{+}=L^{-2}/2$ (i.e., an average of 1/2 of active
site in the system, due to averaging over up-spin and down-spin ground
states) is reached. One large collection of positive excess is moved
around the system in samples with positive magnetization. This packet
of excess annihilates broadly distributed sinks having a magnitude
approximately proportional to $\Delta$. The density of negative-excess
sites (sinks) decreases very rapidly, though the rate of decrease
slows somewhat after the coalescence of positive excess into a single
packet.}
\end{figure}

We also studied how densities of sinks and active sites converge to
their final values $N^{-}L^{-d}$ and $N^{+}L^{-d}$. We have computed
the number densities $\rho^{\pm}(\nPR,L)$ of the positive and negative
sites as a function of {}``time'' $\nPR$ (number of PR cycles completed
so far; $\NPR=\nPR$ at the end of the algorithm). The data, plotted
for $d=2$ in Fig.~\ref{cap:DensityVsTime2D}, shows that the number
density of sites with positive excess reaches a plateau consistent
with constant $\rho^{+}=L^{-d}/2$ in the larger samples, i.e., $N^{+}\approx1$
in samples with positive magnetization. In the late-time regime of
few packets (that is, few active sites), the positive excess packets
may move across the sample many times between global relabellings,
in $d=2,3$. The density of negative excess sites decreases rapidly
with the number of PR operations per site, though more slowly after
the positive excess has coalesced.

\subsection{\label{sub:Spatial-structure-of}Spatial structure of the remnant
sinks}

We can study the topography of the $u_{i}$ in the half of the ground
states that have negative magnetization at small $\Delta$. One approach
is to study the probability distribution and correlations of $u_{i}$
at all sites. We carry this out for $d=1$. Very closely related information
can be found from the locations of the sites where $u_{i}=0$ at the
termination of the algorithm. The spatial distribution of these remnant
sinks reflects the history of the cancellation process. We have examined
this spatial distribution in $d=1,2,3$. The terminal height field
$u_{i}$ can also be computed from the final sink locations alone,
when $\Delta\rightarrow0$, so these two descriptions are closely
related. One method to study the distribution of the sinks is to coarse-grain
the distribution of the remnant sinks over a length $b$. The computed
quantity we used, $N(b)$, is defined as the number of non-overlapping
boxes of dimension $b^{d}$ that contain at least one sink. For a
pure fractal, this would be used to estimate the box-counting dimension
of the set. (One other characteristic of the final topography, which
we study in Sec.~\ref{sec:General-disorder-and}, is the geometry
of paths along unsaturated bonds that start from sinks: these are
the paths followed in global updates to identify reachable sites.
At small disorder, these paths have trivial geometry.)

\subsubsection{Remnant sinks for $d=1$}

\begin{figure}
\includegraphics[%
  width=1.0\columnwidth]{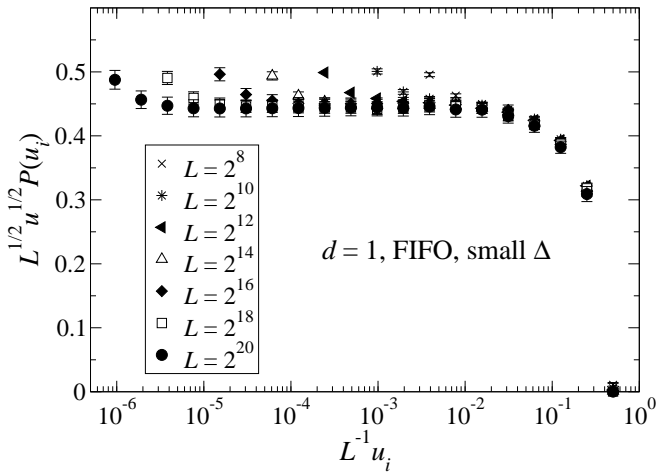}

\caption{\label{cap:P_of_u_i__one_d}Plot of the scaled height distribution
$L^{1/2}u^{1/2}P(u)$ vs.~the sample-size normalized height, $L^{-1}u$,
measured at the termination of the push-relabel algorithm for one-dimensional
samples in the limit of weak disorder. The data collapse well to a
constant, indicating that the distribution of heights behaves as $P(u)\sim u^{-1/2}$
from small scales up to heights $u\approx10^{-2}L$, to within the numerical
error bars. The error bars represent $1\sigma$ estimates for the
sample-to-sample variation; the errors at a given $L$ are correlated,
as the distribution at each scaled distance was computed from a single
set of samples (the number of samples was at least $1700$ at each
sample size). The global update frequency was set at $\Gamma=2L$.}
\end{figure}

After the terminal global update, the sites $i$ that are not sinks have
a height label $u_{i}$ that equals the lattice distances to the nearest
sink. Fig.~\ref{cap:P_of_u_i__one_d} displays a plot of the scaled
height probability distribution $L^{1/2}u^{1/2}P(u)$, where $P(u)$
is defined as the probability of a site $i$ to have a height $u$,
for samples of various sizes $L$ at small $\Delta$. The plots show
that $P(u)$ is well fit by a single power-law behavior, $P(u)\sim u^{-\tau}$,
with $\tau=0.500\pm0.005$. (This very small error bar in $\tau$
is estimated by finding the range of values which give a plateau to
within statistical error for $P(u)L^{\tau}$, for $u>100$
and $2\,097\,152\ge L\ge16\,384$ and assuming that the corrections
to scaling are very small, so that deviations from a plateau represent
only a statistical error in the exponent.)

\begin{figure}
\includegraphics[%
  scale=0.88]{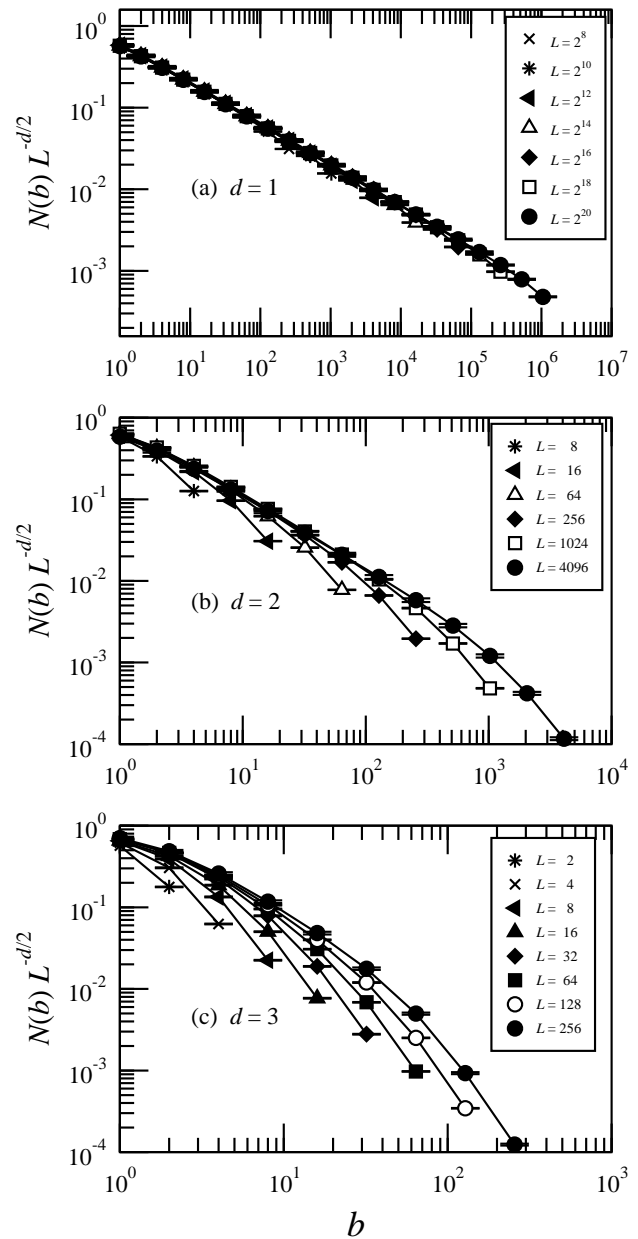}

\caption{\label{cap:BoxCount}Plot of normalized box-counting data $N(b)L^{-d/2}$
vs.~box size $b$ at $\Delta\rightarrow0$. The quantity $N(b)$
counts the number of non-overlapping volumes of linear size $b$ that
intersect the set of remnant sinks. The data shown here is averaged
over samples for the auxiliary field configurations computed at the
termination of the standard (non-blocking) push-relabel algorithm. }
\end{figure}

We also characterized the sink distribution using $N(b)$. The estimates
for this function are plotted in Fig.~\ref{cap:BoxCount}(a) for
$d=1$. This data is also fit by a single power law, with $N(b)\sim b^{-D}$,
where $D=0.500\pm0.005$ (statistical error only). This estimate for
the fractal dimension is consistent with $D=\frac{1}{2}$ for the
set of remnant sinks in the small disorder limit.

The result for fractal dimension $D$ and height-distribution exponent
$\tau$ are related. In one dimension the distribution function $P(u_{i})$
is precisely related to the number $n_{s}(\ell_{ss})$ of sink-to-sink
gaps of length $\ell_{ss}$ via \begin{equation}
P(u_{i})=n_{s}(2u_{i}-1)+2\sum_{k=u_{i}}^{\infty}\left[n_{s}(2k)+n_{s}(2k+1)\right]\,.\label{eq:HtSep}\end{equation}
 In the continuum height limit, then, taking $\frac{dN(b)}{db}\sim n_{s}$
and $P\sim\int_{u_{i}}^{\infty}n_{s}(\ell_{ss})\, d\ell_{ss}$ gives
$\tau=D$, consistent with our estimated values. 

The structure of the sinks is very suggestive, when one considers
the apparent relationship between the dynamics of the push-relabel
algorithm in the limit of small-disorder and studies of the $A+B\rightarrow\emptyset$
reaction. In particular, Leyvraz and Redner studied the fractal structure
of the $B$ particles in the limit that the $B$ particles are immobile.
The primary difference between their analysis and this model is that
the $A+B$ reaction was considered for diffusive $A$ particles. Here,
the $A$ particles (corresponding to positive excesses) move directly
towards the nearest $B$ particle (negative excess sites).

Our presumption will be that this distinction affects only the relationship
between length scale and times. In the diffusive case, the time scale
$t$ gives a length scale $\ell\sim t^{1/2}$. In the case of the
push-relabel algorithm, we will take $t\sim\ell\ln(\ell)$ or, equivalently
up to logarithms of logarithms, $\ell\sim t(\ln t)^{-1}$. This dependence
comes from the linear {}``attraction'' between the positive and
negative excess sites, with a logarithmic correction reflecting the
changes in direction that take place upon annihilation of positive
and negative excess on successively larger scales, consistent with
the finite-size scaling of the running times.

Given this correspondence, we expect the same domain structure of
the negative excess sites in the final states as for the $B$ sites
in the annihilation reaction with immobile $B$ particles. Leyvraz
and Redner, using random walk arguments to sum up the densities of
$A$ vs.~$B$ particles, find the distribution of the distances $d_{BB}$
between neighboring $B$ sites to have the form $P(d_{BB})\sim d_{BB}^{-3/2}$.
Identifying our $\ell_{ss}$ with $d_{BB}$ gives $\frac{dN(b=\ell_{ss})}{d\ell_{ss}}\sim n_{s}(\ell_{ss})\sim\ell_{ss}^{-3/2}$,
which is numerically consistent with our statistics for the final
configuration from the push-relabel algorithm. It is worthwhile to
note that this type of picture is also in agreement with the structure
of the 1D RFIM ground state reported by Schr\"oder \emph{et al.}
\cite{SchroderKnetterAlavaRieger2002}, which was also derived using
absorbing states of random walks. These connections support a unified
picture of the dynamics of the push-relabel algorithm, the structure
of the RFIM ground state, and the previously distinct study of annihilation
reactions.

\subsubsection{\label{sub:Structure-of-remnant}Structure of remnant sinks, $d>1$}

The results we obtain for small disorder in higher dimension are somewhat
more complex. Scaling behavior is also seen, though the results can
depend on the details of the algorithm. In the small disorder limit,
the paths to the sinks are still linear, i.e., non-fractal and straight
(the paths of course must be linear at all disorders when $d=1$).
However, the fractal structure of the remnant sinks is more apparent
in higher dimensions. In some cases there appears to be a new length
scale, intermediate between the microscopic length and the sample
size, that characterizes the final topography in samples with net
negative magnetization.

\begin{figure}
(a) \includegraphics[%
  width=0.80\columnwidth]{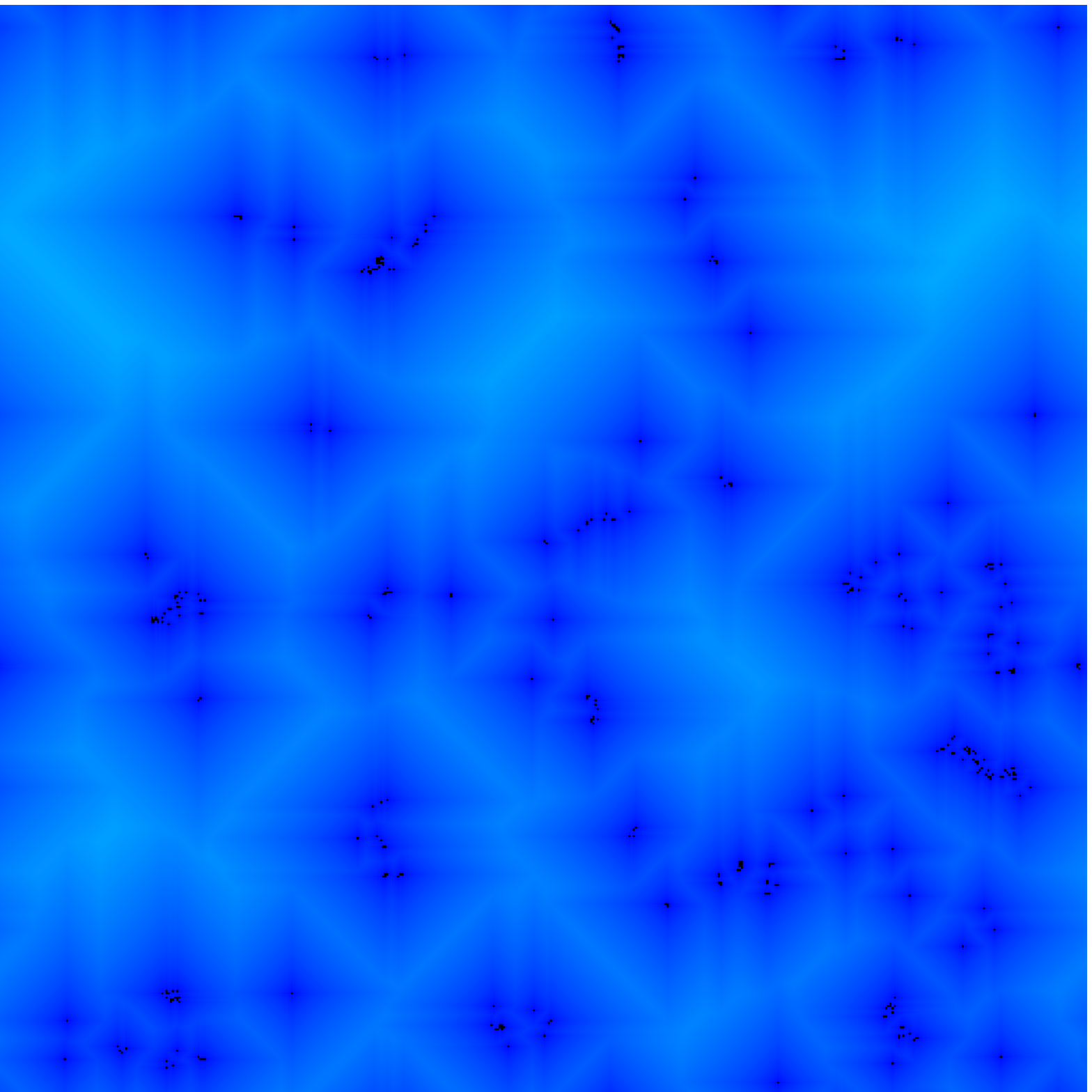}\\

(b) \includegraphics[%
  width=0.80\columnwidth]{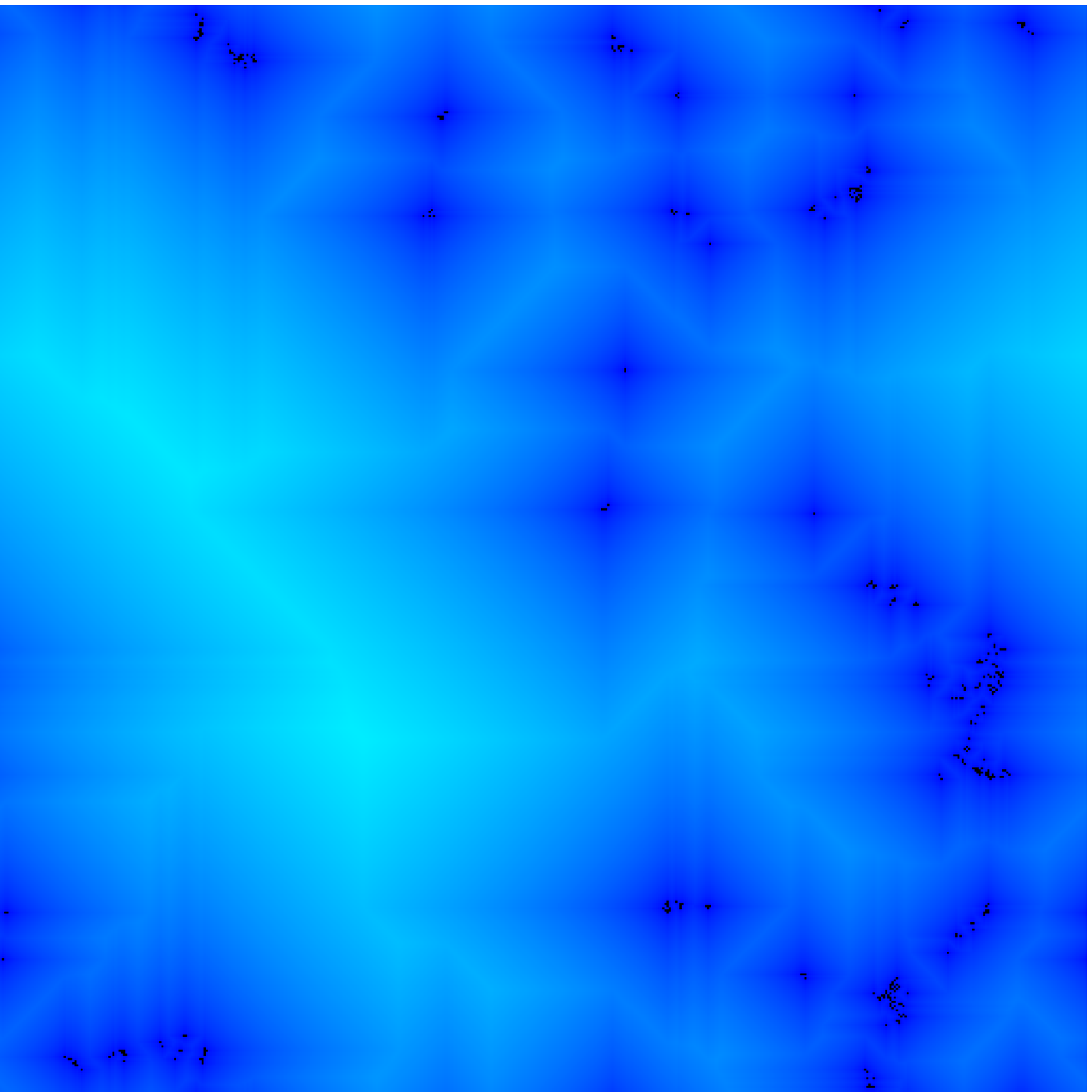}

\caption{\label{cap:TopoPic}{[}Color online{]} Topography of the height field
upon termination of the push-relabel algorithm, for a $d=2$ sample
of linear size $L=512$, with weak disorder ($\Delta\rightarrow0$
limit) for two different modifications of the push-relabel algorithm.
The sample shown has negative magnetization. The lightness of the
color (or grayscale) indicates the magnitude of the height field.
Black pixels indicate spins with negative excess (height $u_{i}=0$),
dark pixels indicate smaller heights, and light regions indicate higher
heights. The height field and remnant sinks are shown for (a) the
push-relabel algorithm with no limits on coalescence and (b) a modification
where positive excess is not allowed to coalesce (the {}``blocking''
version).}
\end{figure}

The topography at the termination of the algorithm is displayed for
$d=2$ in Fig.~\ref{cap:TopoPic} for two variants of the PR algorithm.
One variant is standard: pushes are executed without regard to the
status of the destination site. We also consider in this section a
{}``blocking'' variant of the algorithm that forbids coalescence
of positive excess. Whenever a push is attempted in the blocking version
from a site $i$ with height $u_{i}$ to a site $j$ with $u_{j}=u_{i}-1$,
the algorithm first checks whether the destination site already has
positive excess. If there is positive excess at $j$, the algorithm
does not push excess in that direction, but also does not relabel
site $i$ (a relabel would be executed if there was no push due to
saturated bonds with $r_{ij}=0$). This non-blocking variant slows
down the algorithm some, but prevents the coalescence of the positive
excess into a single packet. The number of positive excess sites does
not decrease as quickly as in the non-blocking version, with the number
of excess packets remaining comparable to the number of sinks, until
$N^{+}$ or $N^{-}$ approach $L^{d/2}$. As can be seen in the figure,
the sites with negative excess are clearly clustered on several scales
for both variants of the algorithm.

\begin{figure}
\includegraphics[%
  width=0.88\columnwidth]{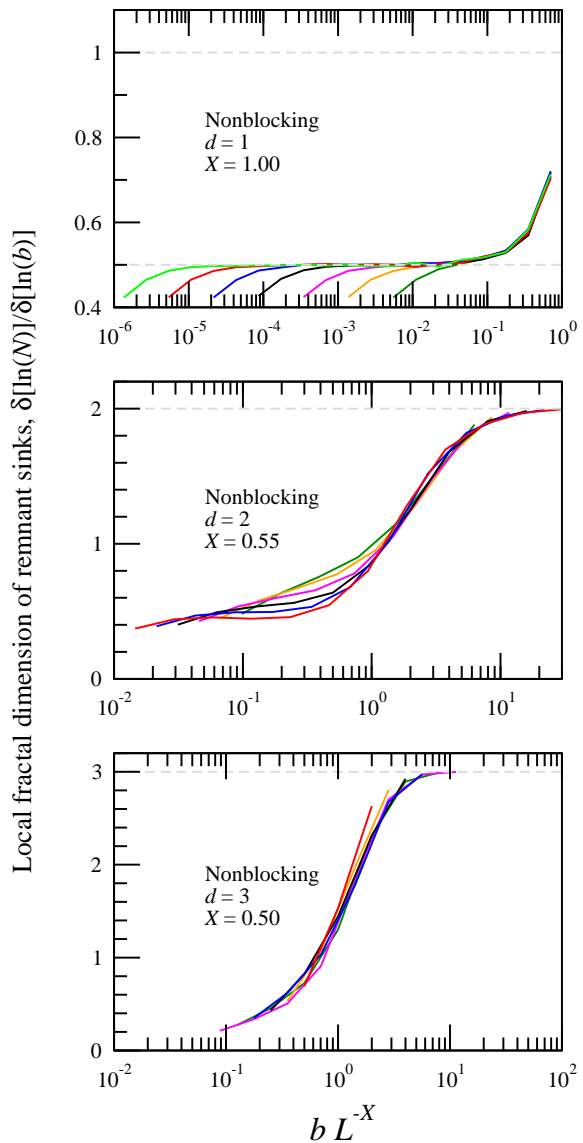}

\caption{\label{cap:box_localderiv_noblock}{[}Color online{]} Plot of the
local dimension -$\frac{\delta[\ln(N)]}{\delta[\ln(b)]}$ of the set
of remnant sinks plotted vs.~scaled box size $bL^{-X}$. The data
is sampled at the termination of the standard (non-blocking variant)
push-relabel algorithm, for $d=1,2,3$. The sample sizes are $L=2^{8},2^{10},\ldots,2^{20}$
from the rightmost to the leftmost curves for $d=1$, $L=2^{7},2^{8},\ldots,2^{12}$
from the highest to the lowest curves for $d=2$, and $L=2^{3},2^{4},\ldots,2^{8}$
for $d=3$ (these curves overlap significantly). In the color version,
the line colors run through the sequence dark green, orange, magenta,
black, blue, red, light green, from smallest to largest $L$. The
dashed light horizontal lines indicate local fractal dimension of
$D=d$.}
\end{figure}

We studied this clustering for both variants by again applying box
counting for the sites with negative excess. The box-counting data
for the non-blocking variant, with $N(b)$ normalized by the number
of sinks $N(0)\sim L^{d/2}$, is plotted as $L^{-d/2}N(b)$ vs.~box
size $b$ in Fig.~\ref{cap:BoxCount}(b,c). We examined this data
for scaling behavior. The error bars are small enough that plotting
the local exponent or slope of the $\ln$-$\ln$ plot was useful in
studying the scaling. The scaling ansatz is that there is a crossover
in the scale-dependent fractal dimension $D(b)$ at a scale $L^{X}$,\begin{equation}
D(b)=\tilde{D}(bL^{-X})\,,\label{eq:Dscale}\end{equation}
with $\tilde{D}(z)$ constant at small $z$ and crossing over to $\tilde{D}=d$
at large argument. The quantity used to estimate $D(b)$ is the discretized
logarithmic derivative, $\delta\ln[N(b)]/\delta\ln(b)\equiv\left\{ \ln[N(2^{1/2}b)]-\ln[N(2^{-1/2}b)]\right\} /\ln(2)$,
whose negative gives an effective dimension when plotted as a function
of $b$. The scaled plot of this estimate for $D(b)$ vs.~$L^{-X}b$
is shown in Fig.~\ref{cap:box_localderiv_noblock}, for our best
fit values $X\approx1.0$ ($d=1$), $X\approx0.55$ ($d=2$) and $X\approx0.5$
($d=3$). Our estimates for the systematic error bars for $X$ are
somewhat smaller for $d=1$ (less than 0.08) than for $d=2,3$, where
variations in $X$ of the order of 0.12 provide plausible, but less
clean, collapse to a single curve for $\tilde{D}$. The values of
$D(b)$ or $\tilde{D}(b)$ at small argument provide an estimate for
the effective box-counting dimension of the sinks at small scales.
The $d=1$ data, as discussed in the previous section, give a fractal
dimension consistent with $D=\frac{1}{2}$ at small scales. The $d=2$
data are consistent with a convergence to an effective dimension of
$D\approx0.4\pm0.1$ over about 1 decade in scale at the smaller scales
for the largest samples ($4096^{2}$) where the longest plateau in
effective dimension is seen. The $d=3$ data are consistent with $D<0.2$
at small scales. At large scales, $b>L^{X}$, the data are quite consistent
with $D=d$, i.e., a scale-independent density of sinks. 

\begin{figure}
\includegraphics[%
  width=0.88\columnwidth]{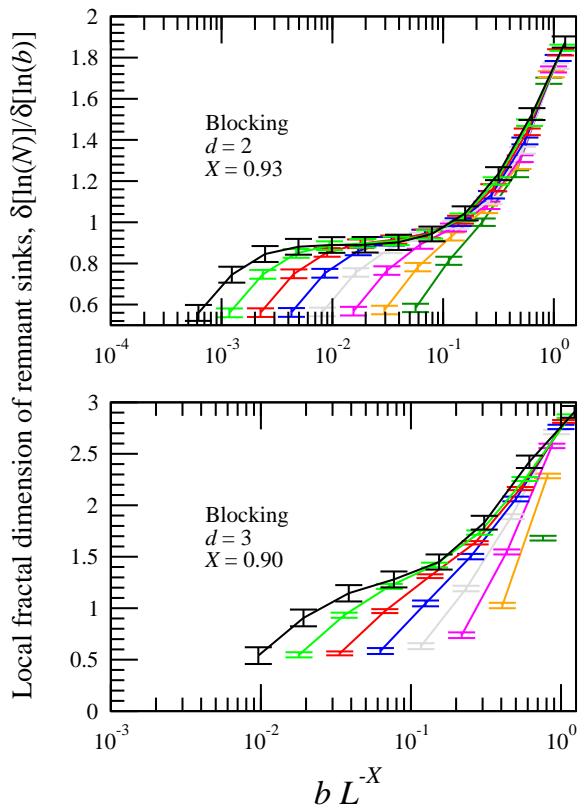}

\caption{\label{cap:box_localderiv_block}{[}Color online{]} Plot of the local
dimension $-\frac{\delta[\ln(N)]}{\delta[\ln(b)]}$ of the set of
remnant sinks vs.~scaled box size $bL^{-X}$ averaged over samples
at the termination of the blocking variant of the push-relabel algorithm,
for $d=2$ and $d=3$. The sample-size ranges are $L=2^{5}$, $2^{6}$,
$\ldots$, $2^{12}$ for $d=2$ and $L=2^{1}$, $2^{2}$, $\ldots$,
$2^{8}$ for $d=3$, with the leftmost curves corresponding to the
largest samples. In the color version, the line colors run through
the sequence dark green, orange, magenta, gray, blue, red, light green,
black, from smallest to largest $L$. The error bars are determined
by $1\sigma$ sample-to-sample fluctuations.}
\end{figure}

The logarithmic derivatives of the box-counting data for the blocking
variant of FIFO in $d=2,3$ are displayed in Fig.~\ref{cap:box_localderiv_block}.
(For the case $d=1$, we find that while the number of active sites
evolves differently, with the assumed form $N^{+}\sim L^{d/2}$ consistent
with the data, the dimensions $D\approx0.5$ and scaling $X\approx1.0$
from the box-counting data are the same for the blocking and non-blocking
variants.) The best collapses are seen for $X$ slightly less than
1, $X=0.93,0.90$ for $d=2,3$, respectively, with an estimated error
of about 0.1. At small scales in $d=2$, the fractal dimension $D$
is approximately $0.9\pm0.1$ at small scales. For $d=3$, the convergence
is less clear, but is consistent with $D$ between 1.0 and 1.5 at
small scales. Assuming a single fractal dimension over all scales,
except for corrections near $b=L$, i.e., taking $X=1$, would give
$D=d/2$, consistent with our data for the blocking variant.

Comparing the results for the distribution of sinks in these two variants
of the algorithm and the density of packets discussed in Sec.~\ref{sub:Number-of-remnant}
leads to a consistent qualitative picture for the interacting evolution
of the active sites, sinks, and topography. In the non-blocking variant
of the push-relabel algorithm, it appears that a scale is frozen in
once the positive excess coalesces. The packet containing the positive
excess moves all about the sample, cancelling each small negative
excess site it encounters and changing directions to follow the height
field $u_{i}$, until it is exhausted. This cancellation acts at small
scales. The non-blocking variant, however, maintains roughly equal
numbers of positive excess packets and negative excess sinks, so that
the positive excess packets have a length scale that grows as the
length scale of the sinks does, throughout the algorithm. The length
scale for the topography of the height field is not set by the {}``freezing-out''
of the positive excess into a single packet, then, but by the sample
size. This leads to a scaling for the final distribution of sinks
with $X\approx1$, consistent within the large systematic error bars
for $X$ in $d=2,3$. The non-blocking variant is more consistent
in microscopic rules and with results for the annihilation dynamics
of particles of constant charge.

\section{\label{sec:General-disorder-and}General disorder and critical slowing
down}

We now consider larger values of $\Delta$, where the bonds can be
saturated and block flow, making the rearrangements of excess more
complicated than the $\Delta\rightarrow0$ limit. We focus on the
true transition at $\Delta=\Delta_{c}$ in $d=3$, but have also studied
the topography of $u_{i}$ and timing for $\Delta\gtsimeq\Delta_{x}$
in $d=1,2$. Ogielski noted \cite{Ogielski1986} that the PR algorithm
takes more time to find the ground state near the transition in three
dimensions from the ferromagnetic to paramagnetic phase. For the highest
priority queue (HPQ) algorithm, identical to LPQ except that active
sites with the highest height are kept at the front of the list, Middleton
and Fisher \cite{MiddletonFisherRFIM,MiddletonCritSlow2002} studied
this critical slowing down more extensively in the case of $d=3$
and work has also been carried out for $d=4$ \cite{Middleton4D}.
Here, we compare the scalings for the timing of FIFO and LPQ algorithms.
The position of the peak in running time can itself be used to estimate
$\Delta_{c}$, if one makes the natural assumption that the critical
slowing down is greatest exactly at $\Delta_{c}$. This critical slowing
down is certainly reminiscent of the slowing down seen in local algorithms
for statistical mechanics at finite temperature, such as Metropolis,
and even for cluster algorithms \cite{SwendsenWang1987,Middleton2004}.

Critical slowing down results from the long length scales that arise
near a continuous phase transition. To connect the topography of the
$u_{i}$ to the physical system, we conjecture that the scale for
the heights $u_{i}$ is given by the correlation length $\xi$. This
assumption is natural: the maximal height in a domain of linear size
$\xi$ must have a scale $\sim\xi$ for excess to be transported across
a domain. The spin-spin correlation functions die off rapidly over
longer ranges, so that the ground-state computation need not rearrange
excess over scales greater than $\xi$. This conjectured relationship
fits the numerical data well.

\subsection{Timing for $d=1,2$}

In one- and two-dimensional systems, there is a system-size-dependent
crossover from large $\Delta$ to small $\Delta$ behavior. Above
this crossover scale $\Delta_{x}(L)$, physical quantities such as
correlation length and algorithmic quantities such as $\NPR/n$ are
expected to converge at large enough $L$ to an infinite-volume limit.
This limit is expected to exhibit critical divergences with a critical
point at $\Delta=0$.

\begin{figure}
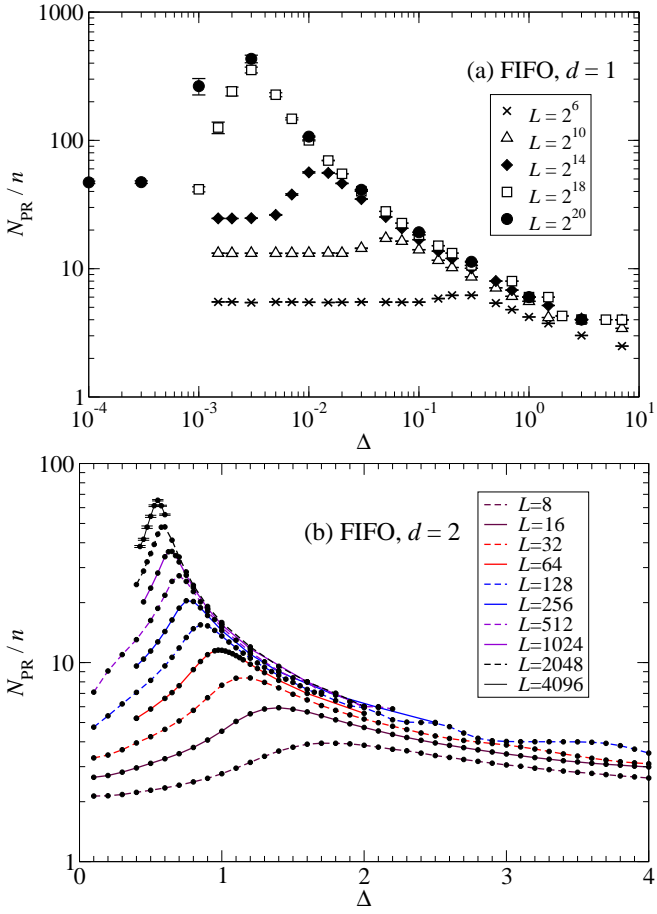

\includegraphics[%
  width=1.0\columnwidth]{Fig10a.eps}

\includegraphics[%
  width=1.0\columnwidth]{Fig10b.eps}

\caption{\label{cap:NPR_1_2}Running times for FIFO algorithm, (a) $d=1$
and (b) {[}color online{]} $d=2$ as a function of disorder $\Delta$
and normalized by sample volume. The error bars are small for $d=2$,
so they are shown only for the largest size, where the error bars
are largest.}
\end{figure}

Our unscaled data for the running time in $d=1$ and $d=2$ for FIFO
are plotted in Fig.~\ref{cap:NPR_1_2}. From the known scaling for
the correlation length $\xi\sim\Delta^{-2}$ \cite{Imry1975,Nattermann1988,SchroderKnetterAlavaRieger2002}
and taking $\NPR/n\propto\xi\ln\xi$, one expects a straight line
fit for $\NPR/n$ when plotted vs.~$\Delta^{-2}\ln\Delta$. We do
not find convergence to a single exponent, but instead effective power
law ranges from $\NPR/n\sim(\Delta^{-2}\ln\Delta)^{0.25}$ to $\NPR/n\sim(\Delta^{-2}\ln\Delta)^{0.7}$
over the range $\Delta=0.02$ to $\Delta=0.005$ for $L=2^{24}$.
It may be that even larger samples are needed to see the expected
divergence in $\NPR$. Despite this, over the range $L=2^{14}\rightarrow2^{24}$,
we do find that the location of the peak in the running time is consistent
with the scaling $\Delta_{x}\sim L^{-1/2}$. The peak in the running
time per site therefore does occur when $\xi\sim L$.

\begin{figure}
\includegraphics[%
  width=1.0\columnwidth]{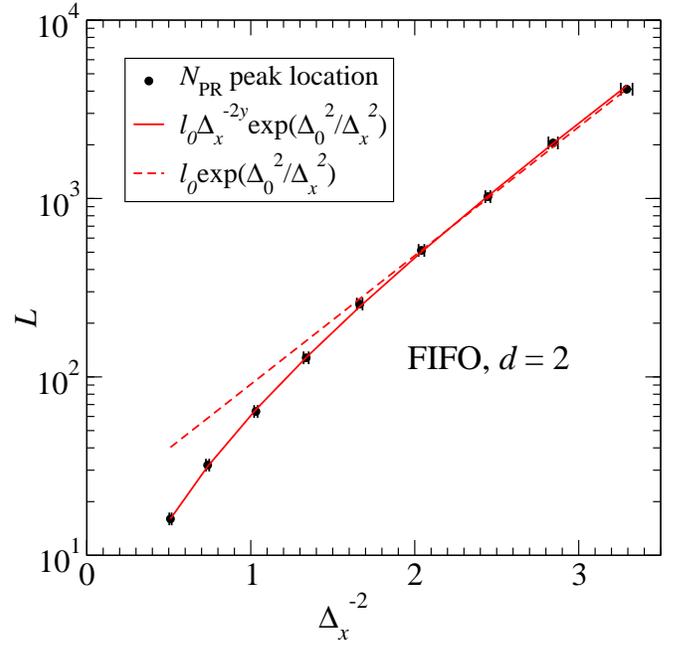}

\caption{\label{cap:2Dpeakloc}{[}Color online{]} A plot of the system size
$L$ vs.~$\Delta_{x}^{-2}$ (circles), where $\Delta_{x}$ is the
fit to the locations of the peaks in $\NPR$ displayed in Fig.~\ref{cap:NPR_1_2}.
The dashed lines show a fit to the simplest expected form $L\sim\exp\left(-\Delta_{0}^{2}/\Delta_{x}\right)$
for $L\ge256$ while the solid line is a fit to all data ($L\ge16$)
using the form $L\sim\Delta_{x}^{-2y}\exp(-\Delta_{0}^{2}/\Delta^{2})$,
with best fit values $\ell_{0}=17.3$, $y=1.07$, and $\Delta_{0}^{2}=1.27$.}
\end{figure}

A similar situation exists in $d=2$: the divergence of the running
time with $\Delta$ is not fit well by the simplest scaling expectations.
However, the location of the peak in the running time does scale in
the expected fashion. The fitted location of the peak for the running
time $L^{-2}\NPR(\Delta)$ is plotted in Fig.~\ref{cap:2Dpeakloc}.
This data is quite consistent with $L\sim\ell_{0}\Delta^{-2y}e^{\Delta_{0}^{2}/\Delta^{-2}}$
\cite{GGSM}, with best fit values $\ell_{0}=17\pm4$, $\Delta_{0}^{2}=1.3\pm0.2$,
and $y=1.1\pm0.2$, citing $2\sigma$-error bars and using all of
the data points for $L\ge16$. This error estimate is consistent with
the variation that one gets by changing the fit to $L\ge64$. The
error bars are large enough that $y=1$ is certainly an acceptable
value.

\subsection{Topography for $d=1,2$}

\begin{figure}
\includegraphics[%
  width=1.0\columnwidth]{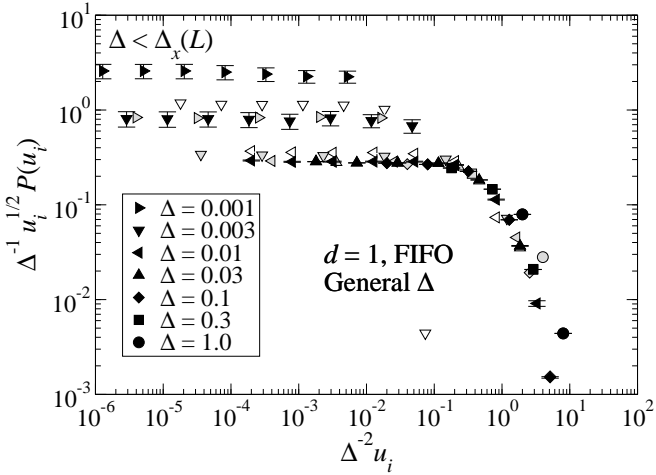}

\caption{\label{cap:Scaling1DhtGenDelta}Scaling plot for the distribution
of the heights at general $\Delta$ for $d=1$ and $L=2^{14},2^{18},$
and $2^{22}$, assuming the scaling form Eq.~(\ref{eq:1DtopoScaling}).
The shape of the symbol indicates the value of $\Delta$, as indicated
by the shape of the solid symbols in the legend. The color of the
symbol indicates $L$, with white-filled symbols for $L=2^{14}$,
gray-filled symbols for $L=2^{18}$, and black-filled symbols for
$L=2^{22}$. The data for a broad range of $\Delta$, including those
values that do not lie in the scaling range $1\gg\Delta\gg\Delta_{x}(L)$
and are therefore not \emph{expected} to collapse, are shown. The
points that do not collapse to the single curve are indicated by the
region labeled $\Delta<\Delta_{x}(L)$ at the top of the graph. The
points that do not fit for well to the right of the collapsed data
are for $\Delta=1$, as indicated by the legend, where the correlation
length is comparable to the lattice spacing. The data that are in
the scaling range give a good collapse to a single curve, with many
points overlapping to the extent that they are not visible. The scaling
curve is flat over the range $10^{-4}<\Delta^{-2}u_{i}<10^{-1}$.}
\end{figure}

In contrast with the attempted scaling for the timing data, the topographical
data for $d=1$ at larger $\Delta$ exhibit a clear scaling with the
expected power-law behaviors. Fig.~\ref{cap:Scaling1DhtGenDelta}
presents an example of this scaling for the fraction $P_{1}(u,L,\Delta)$
of sites with height $u$ for $d=1$. The plot is of the scaled height
distribution function $\Delta^{-1}u^{1/2}P_{1}(u,L,\Delta)$ vs.~the
height normalized by the correlation length, $u\Delta^{2}$, for various
values of $L$ and $\Delta$. This plot assumes a correlation length
$\xi(\Delta)\sim\Delta^{-2}$, with $\xi(\Delta)\ll L$, that $P(u)\sim u^{-1/2}$,
as it is in the limit of small $\Delta$, and a properly normalized
$P(u)$ ($\int du\, P(u)=1$), which together give a scaling form\begin{equation}
P_{1}(u,L,\Delta)\sim\Delta u^{-1/2}\tilde{P}_{1}(u\Delta^{2})\,,\label{eq:1DtopoScaling}\end{equation}
with $\tilde{P}(z)$ constant for small $z$. The data collapse very
well for the range $1\gg\Delta\gg\Delta_{x}(L)$.

\begin{figure}
\includegraphics[%
  width=1.0\columnwidth]{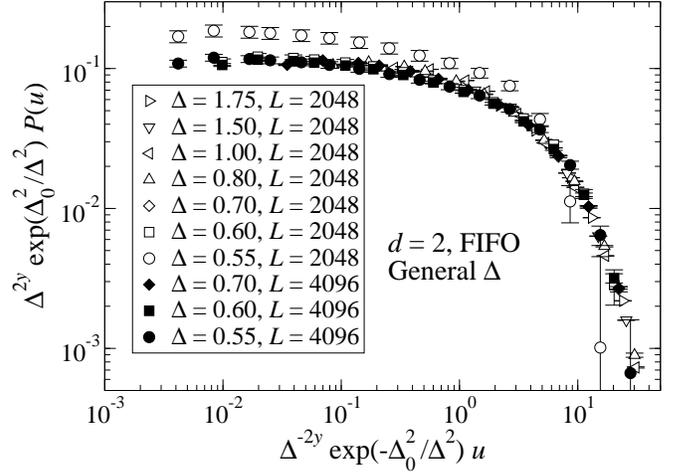}

\caption{\label{cap:HtScale2D}Scaling plot for the distribution $P(u_{i})$
of the final heights $u_{i}$ at general random-field strength $\Delta$
for $d=2$, with $L=2048$ and $L=4096$. The correlation length $\xi$
which sets the scale for the heights $u_{i}$ is taken to diverge
as $\xi\sim\Delta^{-2y}\exp(\Delta_{0}^{2}/\Delta^{2})$, with the
parameters $y=1.07$ and $\Delta_{0}=1.27$ taken from the scaling
for the peak running time (Fig.~\ref{cap:2Dpeakloc}). Curves and
points that are expected to be outside of the scaling range are included
for comparison. Besides small values of $u_{i}$ ($L=2048$ points
near the top-center of the plot), the data that does not collapse
is where$\Delta>1.25$, where $\xi$ is becoming comparable to the
lattice size, and for $L=2048$, $\Delta=0.55$, which is on the low-$\Delta$
side of the running time peak (see Fig.~\ref{cap:NPR_1_2}). The
curves over the range $0.55\le\Delta\le1.00$ collapse relatively
well; by the locations of the finite-size peaks in the running times
of Fig.~\ref{cap:NPR_1_2}, this range of $\Delta$ corresponds to
almost two decades in length scale: $64\le L\le4096$. A fit to the
simpler $\xi\sim\exp(\Delta_{0}^{2}/\Delta^{2})$ with a fit parameter
$\Delta_{0}$ gives a worse scaling collapse.}
\end{figure}

The data for $P_{2}(u)$ also exhibit scaling in $d=2$. The data
then collapse well for disorders $\Delta\gtsimeq\Delta_{x}(L)$, if
we choose a scale for the maximal $u$ that is proportional to $\Delta^{-2y}e^{-\left(\Delta_{0}/\Delta\right)^{2}}\propto\xi$,
where we take the values of $y$ and $\Delta_{0}$ directly from the
fit to the location of the peak in the timing. The scaled data shown
in Fig.~\ref{cap:HtScale2D} are therefore in agreement with a scaling
form\[
P_{2}(u,L,\Delta)\sim\xi^{-1}\tilde{P}_{2}(u/\xi)\,.\]
A fit assuming $\xi\sim e^{\Delta_{0}^{2}/\Delta^{2}}$ gave a somewhat
worse scaling collapse.

\subsection{Timing for $d=3$}

Our data for the running time $\NPR$ lend themselves well to scaling
about the critical point. We have not attempted to separately infer
the critical value $\Delta_{c}$ and the correlation length exponent
$\nu$, which determine the correlation length $\xi$ via\begin{equation}
\xi\sim|\Delta-\Delta_{c}|^{-\nu}\,,\label{eq:xi}\end{equation}
but instead use the values determined in Ref.~\cite{MiddletonFisherRFIM}.
These values, $\Delta_{c}=2.270(5)$ and $\nu=1.37(4)$ were found
from scaling of the stiffness (energy change due to a change in boundary
conditions) and spin-spin correlation functions and are consistent
with, e.g., the location of the peak in the specific heat.

\begin{figure}
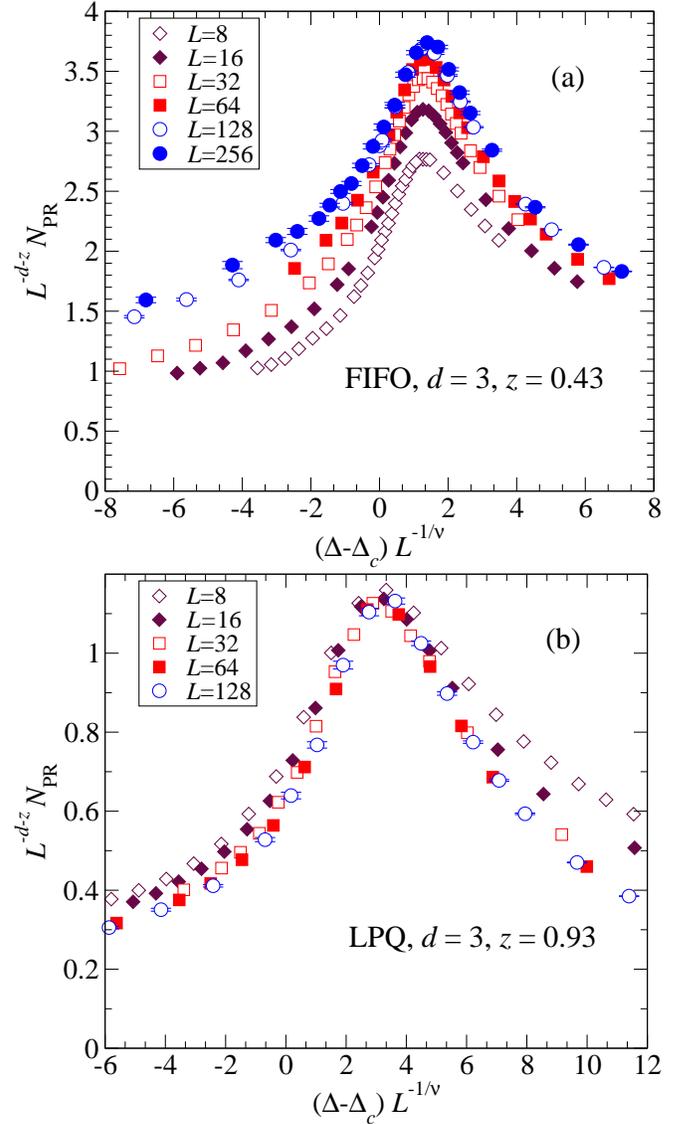

\includegraphics[%
  width=1.0\columnwidth]{Fig14a.eps}

\includegraphics[%
  width=1.0\columnwidth]{Fig14b.eps}

\caption{\label{cap:3DscaledFIFO}{[}Color online{]} Plot of scaled running
time in $d=3$ for (a) the FIFO data structure, with a fit value of
$z=0.43$, and (b) the LPQ data structure, with dynamic exponent $z=0.93$.
The scaling fit assumes $\Delta_{c}=2.270$ and $\nu=1.37$ as fixed
parameters.}
\end{figure}

There is then one parameter to fit, the dynamic exponent $z$, which
describes the divergence in the running time at $\Delta_{c}$, if
one assumes the scaling\begin{equation}
L^{-d}\NPR\sim L^{z}w\left[\left(\Delta-\Delta_{c}\right)^{-1/\nu}L\right]\label{eq:3DNPRscale}\end{equation}
for the number of PR steps per site, where $w(x)\sim x^{-z}$ at large
$x$ and $w(x)\sim|x|^{-z}\ln(|x|)$ as $x\rightarrow-\infty$, to
be consistent with convergence to constant $L^{-d}\NPR$ at $\Delta>\Delta_{c}$
and $L^{-d}\NPR\sim\ln(L)$ for small $\Delta$ (note that the coefficient
of this logarithm in the limit of large $L$ is probably different
from what we find here, given that the true logarithmic behavior may
not have been reached, as discussed in Sec.~\ref{sub:TimeSmallDelta123}).
Our best fits to this scaling form are plotted in Fig.~\ref{cap:3DscaledFIFO}
for the LPQ and FIFO data structures. Our estimates for the dynamic
critical exponent $z$ are distinct for these two structures, with
\begin{equation}
z_{{\rm LPQ}}=0.93\pm0.06;\, z_{{\rm FIFO}}=0.43\pm0.06\,.\label{eq:z}\end{equation}
The error estimates reflect the range of fits that are consistent
with a correction to scaling that is not too large; the statistical
error bars are quite small for this data. The convergence for the
FIFO data may be slower as the data diverge with $L$ more slowly,
so that corrections to scaling arising from, for example, constants
are more evident.

Given the scaling in the data about the thermodynamic critical point,
it is natural to attempt to explain the critical slowing down as being
due to the divergence of a correlation length $\xi$. The finite-size
effect (scaling with $L$) then reflects how the running time diverges
with $\xi$ in the infinite volume limit. The difference in the scaling
of the running times for the two different data structures indicates
different scaling with respect to $\xi$. This difference reflects
the order in which the operations are carried out. For the LPQ algorithm,
only a subset of the sites, those with the lowest $u_{i}$, are subject
to PR operations. This is in contrast with FIFO, where all active
sites are considered cyclically. The FIFO structure leads to a more
rapid coalescence of active sites and cancellation of positive packets
with sinks. It appears that it then takes fewer PR operations to transport
the same quantity of excess across a domain as the domain size increases.
In the LPQ implementation, the value $z\approx1.0$ suggests that
order $\xi$ operations are carried out on average in domains of size
$\xi^{d}$ (note that $z\approx1.0$ in 3D for highest priority queues
).\cite{MiddletonFisherRFIM,Middleton4D}. This is consistent with
each site being relabeled an average of a multiple of $\xi$ times. The LPQ algorithm
coalesces the positive excess by sweeping across the height. The FIFO
algorithm must relabel the sites in such a way that only a small fraction
of the sites are relabeled to height $u_{i}\sim\xi$, with $P(u)\sim u^{z-2}$
if $P(u)$ is described by a simple power law and the number of relabels is
proportional to $\NPR$.
Nonetheless,
taking $\xi$ as a cutoff for the distribution of $u_{i}$ in FIFO
is consistent with the assumption that some portion of the excess
must be rearranged across the width of a domain.

\begin{figure}
\includegraphics[%
  width=1.0\columnwidth]{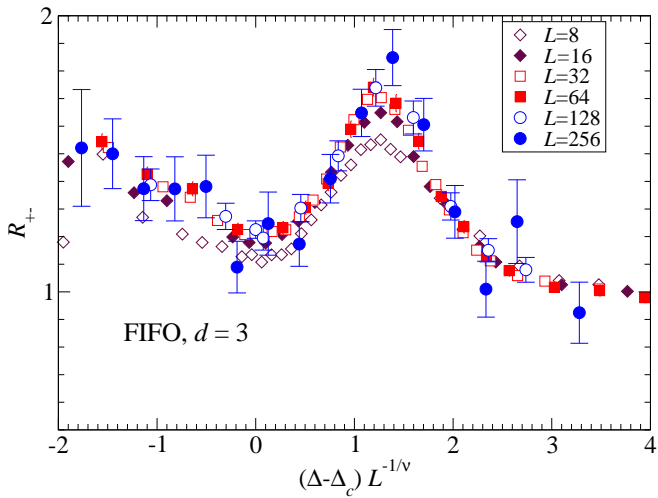}

\caption{\label{cap:Plot-of-ratio}{[}Color online{]} Plot of the dimensionless
ratio $R_{+-}$ of the rms fluctuations $\sigma_{\NPR}^{+}$ in $\NPR$
(time to find the ground state) positive magnetization samples to
the corresponding quantity $\sigma_{\NPR}^{-}$ in negative magnetization
samples plotted vs.~the scaled disorder $(\Delta-\Delta_{c})L^{1/\nu}$,
for samples of sizes $L=8$, 16, 32, 64, 128, and 256. The values
$\Delta_{c}=2.270$ and $\nu=1.37$ were assumed, based on scaling
of physical (not algorithmic) quantities, so this parameter-free fit
is a direct check of scaling for the distributions of running times.}
\end{figure}

We note that there are many distinct scaling plots that can be made
from the timing data. The running times for positively magnetized
and negatively magnetized samples differs significantly for small
$\Delta$, due to the asymmetry between the negative excess sinks
and the positive excess packets at active sites. For large $\Delta$,
the algorithm regains its symmetry, in that the average magnetization
is near zero and any sample is a nearly even mix of up-spin and down-spin
domains. One of the plots that we found to be a sensitive test of
scaling was to compare the running time $\NPR$ between samples with
opposite magnetizations. In particular, the relative \emph{rms fluctuations}
$\sigma_{\NPR}^{+}$and $\sigma_{\NPR}^{-}$ in the operation count
for samples with positive and negative magnetization, respectively,
vary in a rapid fashion near criticality. The dimensionless ratio
$R_{+-}=\sigma_{\NPR}^{+}/\sigma_{\NPR}^{-}$ is a non-monotonic function
characterizing the distribution of running times. This function of
$\Delta$ and $L$ collapses very well for $L\gtsimeq16$ with no
adjustable parameters (taking $\Delta_{c}$ and $\nu$ fixed as stated
earlier). A plot of this collapse is shown in Fig.~\ref{cap:Plot-of-ratio}
for the FIFO algorithm. The parameter-free fit (subject to using published
values for $\Delta_{c}$ and $\nu$) provides a clear confirmation
of scaling for the distribution of running times.

\subsection{Topography at $\Delta=\Delta_{c}$ for $d=3$}

The evolution of the auxiliary fields are closely connected with the
timing of the algorithm, as seen in Sec.~\ref{sec:Small-disorder-limit}
for $\Delta\rightarrow0$. We have studied the final topography of
the height field and the paths that lead to the sinks for $d=3$ and
general $\Delta$. We use this information to study the characteristic
topography at $\Delta$ near $\Delta_{c}$ when using the FIFO data
structure.

A study of the box-counting data for the sinks gives less information
than for $\Delta=0$. For any $\Delta<\Delta_{c}$, there will be
a finite density of minority spins (spins opposite to the net magnetization).
This density drops very rapidly in $d=3$ as $\Delta$ decreases below
$\Delta_{c}$, but these minority spins result in a fractal dimension
of $D=3$ at all scales $\ell$ greater than the typical separation
between the minority spins. We found no obvious singularity in $N(b)$
for $\Delta\approx\Delta_{c}$: the logarithmic derivative for $N(b)$
converges rapidly to $\approx3.0$ for $b>8$, for the disorder range
$2\le\Delta\le3$ . The spatial distribution of the sinks is non-fractal
for $\Delta\approx\Delta_{c}$ (at least from simplest the box-counting
perspective; there will of course be some singularities in the density
and correlations of down spins at $\Delta_{c}$).

\begin{figure}
\includegraphics[%
  width=1.0\columnwidth]{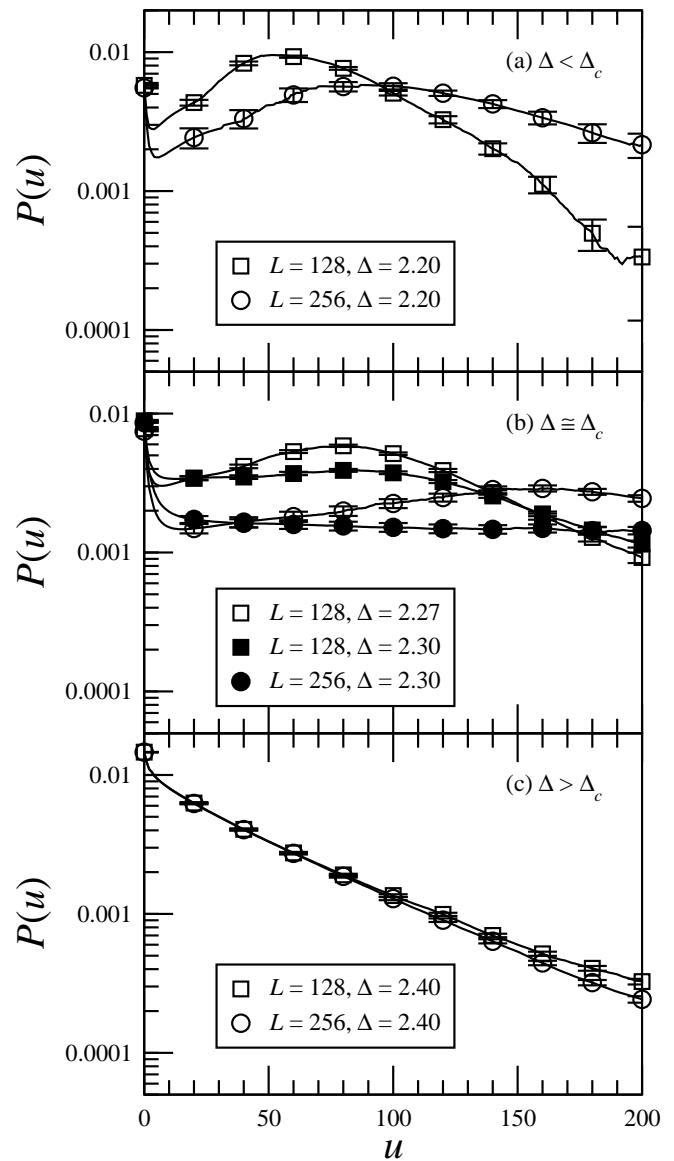}

\caption{\label{cap:3Du}Plot of the height probability distribution $P(u)$
for different values of disorder $\Delta$ and size $L$, for $d=3$
near the critical value $\Delta_{c}=2.27$. The symbols and error
bars are for a sampling of points; the solid lines are measured for
every integer value of $u$. (a) At low $\Delta<\Delta_{c}$, the
distribution of heights $u$ has a peak at an $L$-dependent scale.
(b) Near $\Delta_{c}$, the distribution of $u$ is much less dependent
on $u$ and is nearly constant over a large range of $u$ (on
this scale, power laws are nearly constant), though
the form of $P(u)$ varies rapidly for $\Delta\approx\Delta_{c}$.
(c)High $\Delta$values, here $\Delta=2.40>\Delta_{c}$ give a distribution
$P(u)\sim e^{-u/\xi}$. }
\end{figure}

A feature of the topography that is much more sensitive to the phase
transition is the distribution of the $u$. The height values $u_{i}$
reflect long range correlations of the spins (minority sinks give
$D=3$ but are localized in their effect on the $u_{i}$). In Fig.~\ref{cap:3Du},
we compare the distributions $P(u)$ for the two largest systems we
studied at finite $\Delta$, $L=128$ and $L=256$ to indicate how
$L$ and $\Delta$ appear to affect the height distributions. At large
$\Delta$, the height distribution decreases exponentially with $\Delta$,
$P(u)\sim e^{-u/\xi(\Delta)}$. This is consistent with spin-spin
correlations decreasing exponentially over a characteristic length
scale $\xi(\Delta)$ in the paramagnetic phase. For $\Delta$ significantly
above $\Delta_{c}$, the $P(u)$ are relatively independent of $L$.
For $\Delta\ll\Delta_{c}$, the height distribution peaks at a scale
$u_{i}^{p}\sim L^{X}$ that grows with system size. This is consistent
with the data of Fig.~\ref{cap:box_localderiv_noblock}, which gives
a crossover in the spatial distribution of the sinks (for the non-blocking
variant). Above this length scale $\sim L^{0.5}$, the distribution
of sinks becomes uniform, leading to a cutoff in the $u_{i}$, or
distance to the nearest sink, above that scale. Near criticality,
the distribution $P(u)$ must crossover between these two behaviors,
one decreasing at small $u$ and the other increasing. The distribution
changes rapidly with $\Delta$ and $L$ in this region. We find that
the critical distribution is not varying exponentially rapidly for some $\Delta\approx\Delta_{c}$.

One other feature of the topography that we have investigated is the
structure of the paths that connect down-spin sites to a sink at the
final step of the algorithm. When $\Delta$ is small, these paths
must be linear, as there are no saturated bonds. Near criticality,
many bonds are saturated and there is the potential that the paths
to a sink could be rather torturous, as they would need to avoid saturated
links. Numerical studies were carried out to measure the paths to
the sinks. When carrying out a global update, we maintain a \emph{destination}
\emph{label} that gives the location of the sink to which any positive
excess will flow. For fixed height of each site with $u<\infty$,
the Euclidean distance $r$ of that site to the sink given by the
destination label is computed. The FIFO data can be used to estimate
a fractal dimension $d_{f}$, $u\sim r^{d_{f}}$, for the paths which
guide the pushing of the excess. Near $\Delta_{c}$, it appears that
the paths are nearly linear. For $u>10$ and $\Delta\le2.30$, the
Euclidean distance $r$ was linear in height $u$, with an effective
fractal exponent in the range $0.9<d_{f}<1.1$ for samples of size
$L=256$.

\section{Summary\label{sec:Summary}}

In this paper, we have studied the dynamics of the auxiliary fields
used in the push-relabel algorithm to find the exact ground state
of the random-field Ising magnet, a prototypical glassy model. These
dynamics, and the final state of the fields found during the solution,
reflect both the underlying physical model and the special dynamics
of the algorithm. There is some freedom of choice for these dynamics:
we studied primarily the FIFO queue with global updates, which is
the fastest algorithm we have used and is also most directly similar
to the synchronous evolution used to model physical dynamics. The
evolution of the auxiliary fields is roughly a rearrangement and cancellation
of the locally varying external fields. The extent of this cancellation
determines the size of the domains in the RFIM ground state.

In the limit of small disorder, the dynamics of the field of excesses
is that of a potential-driven annihilation reaction, $A+B\rightarrow\emptyset$.
The algorithmic time per spin to find the ground state for $N$ spins
is apparently linear in $\ln(N)$, though a simple logarithmic behavior
is seen only at large system sizes. The densities of the mobile positive
excess fields (active sites) and the immobile negative excess fields
(sinks) decreases very rapidly with time. When the positive excess
is allowed to coalesce, the number of positive sites becomes of order
unity before the ground state is found. This coalescence leads to
a cutoff in length scale $L^{X}$ in a sample of linear size $L$
that limits the range of scales over which the sinks at the end of
the algorithm can be described as fractal. If coalescence is forbidden,
the remnant sinks appear be described by a fractal measure all the
way up to scale that is nearly equal to $L$. The fractal dimension
in $d=1$, $D\approx0.50$, is consistent with random walk arguments
that have been developed for diffusive annihilation-reaction dynamics
\cite{FLSR92}. For $d=2$, $D\approx0.4$ and the fractal dimension
$D<0.2$ for $d=3$.

Diverging length scales in the RFIM influence the running time for
the algorithm at general disorder values. While the running time itself
is not easily scaled, the peak location for $d=1,2$, $\Delta_{x}(L)$,
scales as expected from calculations for the divergence of the correlation
length $\xi$ with $\Delta$ in the RFIM \cite{ImryMa1975,Binder1983,GrinsteinMa82}.
In particular, for $d=2$, the location of the peak in the running
time for a finite system and the distribution of the height fields
near the transition give evidence for the predicted form for the power
law corrections to the exponential dependence of $\xi$ on $\Delta^{-2}$.
Further examination of the data may be useful in explaining how $\NPR(\Delta)$
diverges. In $d=3$, the scaling of the running time (and its distributions,
as seen in Fig.~\ref{cap:Plot-of-ratio}) are quite consistent with
values \cite{MiddletonFisherRFIM} for the critical value $\Delta_{c}$
and correlation length exponent $\nu$ obtained from simulations for
the physical properties of the RFIM. For $\Delta\approx\Delta_{c}$
in $d=3$, the distribution of the field corresponding to the potential
$u_{i}$ generated by the sinks is nearly constant in $u_{i}$. For
small $\Delta>\Delta_{x}$, $P(u_{i})\propto u_{i}^{-1/2}$ for $u\ll\xi$
in $d=1$ and varies very slowly for $u_{i}$ up to $\xi$ in $d=2$.

We believe that these consistent and strong connections between algorithm
dynamics, the physics of the RFIM, and the mathematics used to describe
annihilation processes provide insight into both the push-relabel
algorithm and the RFIM. One is tempted to speculate that relating
the rearrangement of fields in the push-relabel algorithm to finding
the domains in the RFIM might help in analytic approaches, either
in finite dimensions or on hierarchical lattices \cite{MSC93}. The
running time of the PR algorithm is directly related to the average
of the potential or height field; distinct algorithms give different
scaling for this average, though the physical ground state that is
found is identical. A better understanding of this dynamical exponent
$z$ given by the average height, is likely to shed light on both
the algorithm and the RFIM ground state at criticality.

\begin{acknowledgments}
This work has been supported by the National Science Foundation under
grants ITR DMR-0219292 and DMR-0109164. AAM would like to thank the
Kavli Institute for Theoretical Physics, where part of this work was
carried out, for their hospitality and A.~Rutenberg, J.~Machta,
and B.~Vollmayer-Lee for useful discussions.

\bibliographystyle{apsrev}
\bibliography{./rfim,./cs,./misc_cs_rfim,./dmr04}
\end{acknowledgments}

\end{document}